\documentclass[a4paper,11pt]{article}
\pdfoutput=1 

\usepackage{jcappub} 

\usepackage[T1]{fontenc} 

\newcommand{\CauA}{\mathcal{A}^2(r)}
\newcommand{\gbeta}{g_{\beta}(r)}
\newcommand{\Msun}{ 10^{14} M_{\odot}}
\newcommand{\Mf}{ M_{200}}
\usepackage{rotating}  
 \usepackage{verbatim}
\usepackage{makecell} 

\def\doi{http://doi.org}

\newcommand{\HCd}{\mathcal{H}}

\def\HCdt0{\tilde{\HCd}_{0}}

\usepackage{multirow}
\usepackage{tikz}

\title{\boldmath Unveiling the Coma Cluster Structure: From the Core to the Hubble Flow}

\author[a]{D. Benisty,}
\author[b,c,d]{J. Wagner,}
\author[e,f,g]{S. Haridasu,}
\author[e,f,g]{and P. Salucci}

\affiliation[a]{Leibniz-Institut fur Astrophysik Potsdam (AIP), An der Sternwarte 16, 14482 Potsdam, Germany}
\affiliation[b]{Academia Sinica Institute of Astronomy and Astrophysics, 11F of AS/NTU Astronomy-Mathematics Building, No.1, Sec.~4, Roosevelt Rd, Taipei 106216, Taiwan, R.O.C}
\affiliation[c]{Helsinki Institute of Physics, P.O. Box 64, FI-00014 University of Helsinki, Finland}
\affiliation[d]{Bahamas Advanced Study Institute and Conferences, 4A Ocean Heights, Hill View Circle, Stella Maris, Long Island, The Bahamas}
\affiliation[e]{SISSA-International School for Advanced Studies, Via Bonomea 265, 34136 Trieste, Italy}
\affiliation[f]{IFPU, Institute for Fundamental Physics of the Universe, via Beirut 2, 34151 Trieste, Italy}
\affiliation[g]{INFN, Sezione di Trieste, Via Valerio 2, I-34127 Trieste, Italy}

\emailAdd{benidav@aip.de}
\emailAdd{wagner@asiaa.sinica.edu.tw}
\emailAdd{sharidas@sissa.it}
\emailAdd{salucci@sissa.it}

\abstract{The Coma cluster, embedded in a cosmic filament, is a complex and dynamically active structure in the local Universe. Applying a density-based member selection (\texttt{dbscan}) to data from the Sloan Digital Sky Survey (SDSS), we identify {cluster member galaxies from} its virialised core {out to the} zero-velocity boundary {in the least model-dependent way}. 
{From \texttt{dbscan}, we infer a projected virial radius of $r_{\rm vir} = \left(1.95 \pm 0.12\right)\,h^{-1}~\text{Mpc}$ and projected zero-velocity radius of $r_{\rm ta} \geq 4.87~{h}^{-1}~\mbox{Mpc}$. Assuming that the barycentre of Coma has zero peculiar velocity, its distance from us is $r_\mathrm{c}=(69.959 \pm 0.012_\mathrm{stat}) \, h^{-1}~\text{Mpc}$ determined from the redshifts of 1092 member galaxies.}
Cross-correlating with the Cosmicflows-4 (CF4) catalogue enables a velocity-distance analysis, incorporating radial {cosmology-independent} infall models and redshift-independent distance estimators. {This reveals, for the first time, the Hubble flow surrounding Coma, a first step to investigate the entanglement between Coma's dark matter halo and the dark energy driving the expansion of the surroundings.  
If $v_\mathrm{c}$ is moving with the cosmic expansion, the CF4 distances yield a Hubble constant $H_0 = (73 \pm {1_\mathrm{stat} \pm 7_\mathrm{sys}})~\mbox{km}/\mbox{s}/\mbox{Mpc}$ with a dominating systematic error from different calibrations for the distance moduli. Mass estimates via caustics, the virial theorem, and the Hubble-flow method yield $M = [0.77, 2.0] \times 10^{15}\,h^{-1}\,M_{\odot}$, consistent with prior mass estimates. However, our mass estimates are based on fewer model assumptions in the member selection and require $\sim80\%$ fewer members to attain the same precision. Our approach maps the structure of Coma into {its Hubble flow and shows degeneracies between the Hubble constant, the virial radius, and the total mass only using data and models from the single line-of-sight towards Coma}.}}

\begin{document}
\maketitle
\flushbottom

\section{Introduction}
\label{sec:intro}

Galaxy clusters are the largest gravitationally bound structures in the Universe, containing galaxies, hot gas, and dark matter~\cite{Bykov:2015yxj}. Dark matter, which constitutes the majority of the total mass of a cluster, plays a crucial role in binding these structures together. Although dark matter does not emit, absorb, or reflect light, its presence is inferred through gravitational effects, such as gravitational lensing, where the cluster mass bends light from more distant objects, \cite{bib:Wagner2019,bib:Lin2022}. This invisible component provides the trough in which the galaxies and the hot gas are moving. However, the dynamics of galaxy clusters are not solely governed by dark matter; dark energy also plays a significant role, particularly in the outskirts of galaxy groups and clusters, \cite{Munshi:2003vu,bib:Sandage1986,Karachentsev:2008st,Chernin:2013rpa,Merafina:2014ysa,bib:Pavlidou2014,Benisty:2023vbz}.

The influence of dark energy becomes more pronounced at larger radii, shaping the Hubble flow around the cluster, where its repulsive effect counteracts the gravitational pull of the cluster mass~\cite{Peebles:2002gy}. This interplay between dark matter and dark energy is crucial for understanding the evolution and future of galaxy clusters, in particular the decoupling of bound cosmic structures from the global cosmic expansion~\cite{Morandi:2016cet,Chernin:2013rpa, 2012A&A...541A..84M,Bisnovatyi-Kogan:2013oqa, Merafina:2014ysa,Benisty:2025mfv}.

The Coma cluster (Abell 1656), located about 100 Mpc from us, is one of the richest and most massive galaxy clusters in the nearby Universe. With a mass around $10^{15}~M_{\odot}$, Coma serves as a dominant gravitational attractor in its environment, influencing the motion of galaxies within tens of Mpc~\cite{bib:Zwicky1937, Malavasi:2019jgy}. Despite its prominence, the Hubble flow around Coma has not been systematically studied to the same extent as the one of the Virgo or Fornax cluster. The larger distance and the challenges associated with obtaining accurate and precise distances and velocities for galaxies in the vicinity of Coma have limited previous investigations. However, advances in observational techniques, including more precise distance indicators, now make it possible to explore the Coma-centric Hubble flow in detail.

{This work explores the structure of the Coma cluster from the virial core out to the Hubble flow for the first time}. By analyzing the velocities and distances of galaxies in the Coma region, we characterize the infall and outflow patterns, determine the turnaround, the virial areas and estimate the cluster mass by different approaches. This study not only enhances our understanding of the impact of Coma on its surroundings, but also contributes to the effort to jointly calibrate different probes of cosmic distances and investigate the tensions between the local and global measurements of the Hubble constant \cite{bib:Scolnic2024, Falco:2013zya, Lokas:2003ks, bib:Tully2023}.

The paper is organised as follows: Section~\ref{sec:member_selection} presents the new {cluster member selection which first employs a redshift-space selection and then performs a subsequent density-based clustering on the sky.  Section~\ref{sec:velocity_distance_relation} discusses the necessary preprocessing steps for the redshift-independent distances of the cluster member galaxies that alleviate the large uncertainties. We then show how the radial infall models, usually used to model velocities in the Local Group or neighbouring structures, can also be employed to much farther structures like Coma. Combining the redshift-independent distances and the radial-infall-model velocities, we set up Coma's Hubble diagram from the centre of mass out to its Hubble flow. In contrast to other approaches, the method developed in this work does not make any assumptions about a background cosmology beyond the local, linear Hubble expansion and does only rely on information along the single line of sight towards the Coma cluster}. Section~\ref{sec:mass_det} details the approaches to estimate the mass of the structure. 
Section~\ref{sec:dis} concludes with a synopsis of all findings and an outlook.

\section{Member Selection}
\label{sec:member_selection}

\subsection{Data Retrieval from SDSS DR17}
\label{sec:data_retrieval}
We base our data selection and preprocessing on the coordinates of the Coma cluster in the Simbad database: $\boldsymbol{c}_\mathrm{c} = (\alpha_\mathrm{c}, \delta_\mathrm{c}) = (194.935000^\circ,27.912472^\circ)$ \citep{bib:Abdullah2020} in spectroscopic redshift $z_\mathrm{c}=0.023333 \pm 0.000130$ \citep{bib:Bilton2018}. 
A \texttt{topcat} TAP query of the SDSS DR17 specobj database is performed\footnote{\url{https://datalab.noirlab.edu/tap}} for objects in class \textit{galaxy} within a rectangle of side length $30^\circ$ in $\alpha$ and $\delta$ around $\boldsymbol{c}_\mathrm{c}$.
All entries with \textit{zwarning}~$\ne 0$ or \textit{zwarning\_noqso}~$\ne 0$ are excluded. 
Entries with \textit{zerr}~$<0$ and \textit{zerr}$>0.0005$ are excluded as well. 
The datacube obtained in the Coma cluster field consists of 185,887 galaxies with spectroscopic, heliocentric redshifts\footnote{\url{https://www.sdss4.org/dr17/spectro/spectro_basics/}}. 
Further constraining the dataset to the redshift range $z \in \left[0.0, 0.05 \right]$, the number of galaxies is reduced to 11,137.
In addition, only considering all galaxies within a radius of $10^\circ$ from $\boldsymbol{c}_\mathrm{c}$ on the sky, we arrive at our final set of 5118 Coma member galaxy candidates. 
Since SDSS DR17 has a limiting detection magnitude in the range of 17 to 22.5 for the $i$-band and galaxy spectra are probed in a redshift range $z\in \left[ -0.01, 1.00 \right]$, the member-galaxy candidates are highly complete to these limits, see \cite{bib:Comparat2017} in general. Moreover, Fig.~4 of \cite{bib:Sohn2017} shows a completeness study of Coma, stating that the cluster is highly complete down to $r=17.77$, the spectroscopic flux limit of SDSS.
Their analysis is based on the caustics method, see Section~\ref{sec:caustics}, to identify member galaxies after the method was validated by a cosmological
hydrodynamical simulation. In \cite{bib:Abdullah2020}, a similar approach is pursued to determine the completeness of their ``GalWeight'' method.
Thus, the completeness measure is based on the number of correctly identified member galaxies in the simulation versus the total number of member galaxies as identified by a Friends-of-Friends clustering. 
Since the definition of a cluster, or a structure in general, is fixed by the choice of clustering approach in the simulation, the completeness measures the ability of the caustics or "GalWeight" method to recover the Friends-of-Friends cluster membership assignment. While this evaluation yields a valuable estimate of undetected member galaxies, we refrain from pursuing such a completeness study due to its strong dependence on many assumptions, mostly the non-uniqueness in the definition of a cluster, and the fact that the very well-studied Coma cluster may also be subject to target-selection biases. Instead, we directly compare our Coma member selection and the inferred characteristics of the cluster with existing ones from the literature. In this way, potential biases due to a sparse sampling of the cluster volume and dependencies on the choice of a specific galaxy-member type, like galaxies that host supernovae, can be investigated in a framework with a minimum of necessary model assumptions.

\subsection{Selection along the Line of Sight}
\label{sec:selection_line_of_sight}

{Instead of jointly selecting Coma member galaxies in the full three-dimensional space spanned by redshift and angular position on the sky, we choose to first select suitable candidates in redshift space. This approach allows us to work directly on the observational data instead of converting them into physical coordinates with additional assumptions. In this way, a large amount of field galaxies are sorted out in a very robust, precise, and efficient way. Subsequently applying a clustering algorithm on the sky is faster and less resource-consuming than a three-dimensional version. The latter also has the disadvantage that it either relies on additional assumptions to convert the observables into physical distances or joins redshifts with angles on the sky into a common datacube}. 


%
To separate Coma from other structures along the line of sight, we identify local maxima and minima in a histogram of redshifts up to $z=0.05$.
As redshift bin size, we choose $\Delta z = 0.0005$, which is the maximum allowed $zerr$ in our dataset. 
While the line of sight towards Coma contains a lot of galaxies, up to the observation limit of SDSS DR17, the highest peak in the number of galaxies $n_\mathrm{gal}$ clearly belongs to the Coma cluster. 

For a robust separation, we create a refined histogram with bin size $\Delta z = 0.0000125$, i.e.~one fourth of the original one. Smoothing the resulting histogram via a one-dimensional Gaussian filter with standard deviation $\sigma=4$, we obtain a smoothed curve at the resolution of the original histogram. Local maxima and minima are determined from the smoothed graph with the \texttt{scipy} function \texttt{find}$\_$\texttt{peaks} in its default configuration, comparing neighboring bin counts and returning those bins whose two neighboring direct bins show lower counts. 

\begin{figure}
\centering
\includegraphics[width=0.55\textwidth]{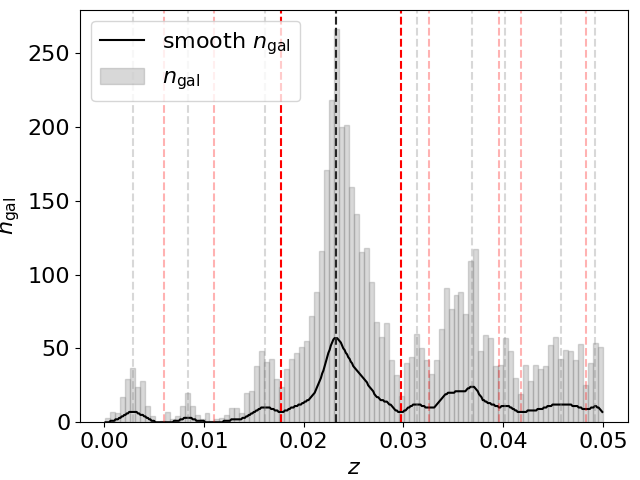}
\caption{\it{Constraining the Coma cluster member galaxies along the line of sight to $z\in \left[0.0177, 0.0297 \right]$ by finding local maxima and minima in a smoothed version (black line) of a redshift histogram (gray bars in the background). Vertical red lines delimit structures according to the minima. The peak position for Coma is found at $z=0.0232$ ($\pm 0.00025$ due to the chosen resolution), indicated by a vertical black line according to the global maximum. Other local maxima are marked with gray vertical lines, denoting peaks of other structures.}}
\label{fig:cluster_separation}
\end{figure}

Fig.~\ref{fig:cluster_separation} shows the resulting partition into clusters along the line of sight based on the smoothed histogram (black line) which is plotted on top of the original histogram (grey bars in the background). As the red vertical lines indicate, the local minima are used to delimit structures, while the grey vertical lines denote the peak positions, i.e.~the local maxima, of these structures.
For the peak corresponding to the Coma cluster, we find its limits as $z\in \left[0.0177, 0.0297\right]$ based on the local minima around the peak at $z=0.0232$.
All these three values have a precision of $\Delta z = 0.00025$, due to the chosen bin size based on the maximum redshift error $zerr$ that can occur. 

{Reducing our datacube to only contain galaxies within the redshift range $z\in [0.0177,$ $0.0297]$, we are left with 2477 Coma member candidates. Since this selection approach does not partition the observed redshift into a cosmic distance along the line of sight and a peculiar velocity on top of the cosmic distance, the selection may be biased due to the Kaiser effect~\cite{bib:Kaiser1987}. Field galaxies with high peculiar velocities may enter the selected redshift range erroneously. Vice versa, actual member galaxies with high peculiar velocities can lie outside the redshift selection and therefore be missed. However, the higher the mean redshift of the cluster, this effect becomes less relevant. For Coma, the total velocity amounts to $cz_{\mathrm{c}}\approx 7000~\mbox{km}/\mbox{s}$, which is 7 times higher than common peculiar velocities in gravitational potentials on a galaxy cluster scale of about $1000~\mbox{km}/\mbox{s}$.
Systematic clustering studies \cite{bib:Knebe2011, bib:Old2014, bib:Old2015} revealed that richness-based clustering algorithms, such as the one that we propose here, suffer from interloper galaxies but still produce the most accurate mass estimates despite biased member selection.} 

\subsection{Selection on the Sky}
\label{sec:selection_sky}

The remaining 2477 Coma member candidates in the suitable redshift range need to be partitioned on the sky into a set of actual Coma members and the field galaxies surrounding Coma. As detailed in the previous section, the pre-selection along the line of sight preserves the full three-dimensional cluster structure except for a small amount of potential interlopers, such that our efficient combination of line-of-sight pre-selection and on-sky clustering captures the shape and members of Coma equally well as three-dimensional clustering approaches.

For this two-dimensional clustering of angular positions on the sky to identify all member galaxies, we choose the Density-Based Spatial Clustering of Applications with Noise, \texttt{dbscan}, of \cite{bib:Ester1996}. The first use of this algorithm in astronomy was in \cite{bib:Turner1976} where it was used to compile the first catalogue of galaxy groups in the local Universe. Its advantages are that it can generate clusters of arbitrary shape and it avoids to create line-like clusters that can often occur in single-linkage methods, {such as Friends-of-Friends}. Important for our one-class-clustering application, it has a notion of outlier points, also called noise, and can therefore recognize clusters surrounded by noise. Consequently, we also circumvent the issue that points at the border between two \texttt{dbscan}-clusters may be assigned to one or the other cluster depending on the order of investigation. \texttt{dbscan}'s average time complexity is only $\mathcal{O}(n \log(n))$, memory storage is $\mathcal{O}(n^2)$ with precomputed distances between the $n$ points in the dataset to be clustered.

As its name implies, \texttt{dbscan} is a density-based algorithm and it employs two parameters to be adapted to the dataset, called \emph{eps} and \emph{min\_samples}.
The former can be understood to create the $\epsilon$-ball around a point and therefore represents the radius of the circle around each data point that needs to contain at least \emph{min\_samples} data points to be assigned to a cluster.
Single-linkage clustering can be considered as a special case, assuming that \emph{min\_samples} = 1, thereby losing the notion of noise points and requiring a post-processing step to eliminate noise clusters containing fewer points than a given threshold.
More details on the mutual relation are given in the review by \cite{bib:Kriegel2011} and references therein.

Even though our approach remains cosmology-agnostic and the two parameters will be fixed based on the structure of the data alone, it is possible to interpret the ratio of \emph{min\_samples} to $\pi(\emph{eps})^2$ as a minimum number density of galaxies (= richness) on the sky that is required for a galaxy cluster. 
Setting \emph{eps} to the virial radius and taking into account the galaxy distribution in redshift, scaling relations like the mass-richness relation can be used to translate the number density of galaxies within the virial radius into a mass density.
A comparison to the virial overdensity $\Delta_\mathrm{coll} = 178$ for virialised structures collapsed in a matter-dominated epoch in the Universe can then serve as a link between data-driven density-based clustering and astrophysical models of structure growth (see, e.g.~\cite{bib:Pace2010} for details on spherical collapse models and the weak dependencies on dark energy which extend the original calculation embedded in a matter-dominated Universe or \cite{bib:More2011} for details how the linkage length is related to the local overdensity of the cluster). 
More directly, we can compare the chosen \emph{eps}-parameter with the linkage length to obtain realistic cosmic structures with a Friends-of-Friends clustering. 
Using the latter, a single-linkage hierarchical clustering, the linkage length is heuristically chosen to be 0.2 times the average inter-particle separation in a cosmological simulation, \cite{bib:Schaller2024} (further refinements to this approach were discussed, for instance, in \cite{bib:Duarte2014}). 
To eliminate noise, the implementation of \cite{bib:Schaller2024} discards structures containing less than 20 points after the Friends-of-Friends clustering.

As distance measure, we choose the great-circle distance between angular positions on the celestial sphere. 
Hence, we do not make a small-angle approximation on the sky, nor do we make any assumptions about the underlying cosmology (by transforming angles into physical distances) in order to keep the algorithm as general and cosmology-agnostic as possible. 

\begin{figure*}
\centering
\includegraphics[width=0.45\linewidth]{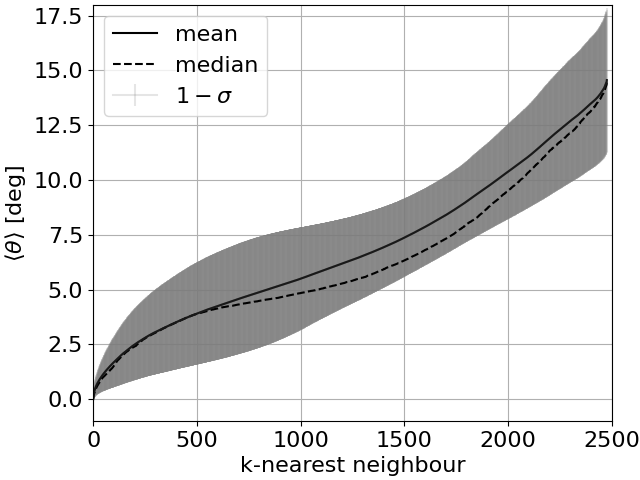}
\includegraphics[width=0.45\linewidth]{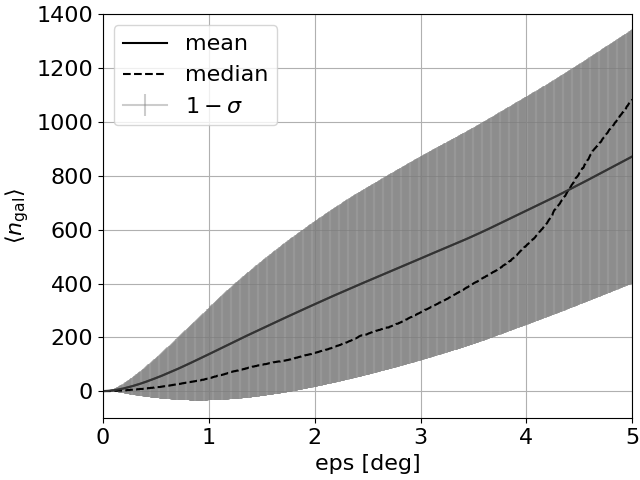}
\caption{\it{Average distance for all $k$-nearest neighbors of all 2744 galaxies in the Coma field (left) and average number of galaxies in a circle of radius \emph{eps} around all galaxies (right). Any $\langle n_\mathrm{gal} \rangle < 0$ is nonphysical but plotted to show the change in the lower 1$-\sigma$ bound in $\emph{eps}\in \left[0,2\right]^\circ$.}}
    \label{fig:dist_mat}
\end{figure*}

From the distance matrix that calculates the great-circle distance between all points, a first estimate for the optimum parameters for \texttt{dbscan} can be constrained. 
First, the mean angular distance on the sky is obtained as $6.9^\circ$, implying a linkage length of $1.4^\circ$ for a Friends-of-Friends clustering. 
This value is about the virial radius which is supposed to lie between $1$-$2^\circ$ on the sky, see, for instance \cite{bib:Sohn2017} with an estimate of $\theta_{200}=1.8^\circ$. 
Next, we determine the average distance, its median, and standard variation for all $k$-nearest neighbours of all data points, as plotted in Fig.~\ref{fig:dist_mat} (left) and the average number of galaxies, its median, and standard deviation in a circle of radius \emph{eps} around each data point in Fig.~\ref{fig:dist_mat} (right). 
The former, in particular the change in the standard deviation, reveals that there is an over-density in the data and it contains around 1000 galaxies. 
Assuming that the over-density causes the median distance to be lower than the mean distance, $\emph{min\_samples}\approx500$ is a good first estimate to identify the cluster. 
Analogously, we can read of Fig.~\ref{fig:dist_mat} (right) $\emph{eps}\approx2^\circ$ for the radius parameter of \texttt{dbscan}. 
Yet, the data-based information cannot provide more than these rough estimates without using models of over-density richness profiles on top of a field of galaxies following a specific stochastic distribution.

\begin{figure*}
\centering
\includegraphics[width=0.325\textwidth]{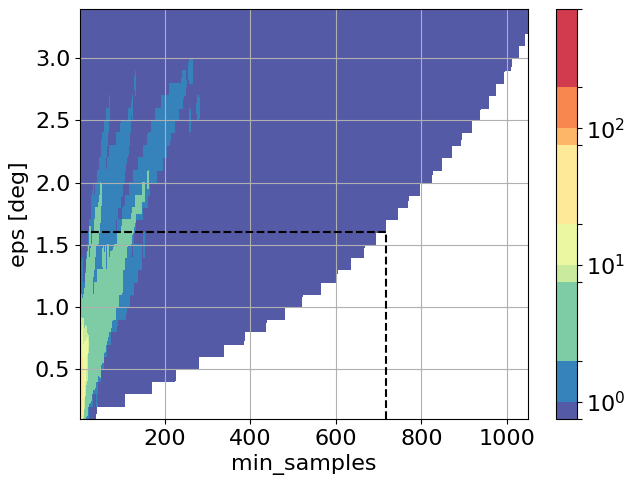}
\includegraphics[width=0.325\textwidth]{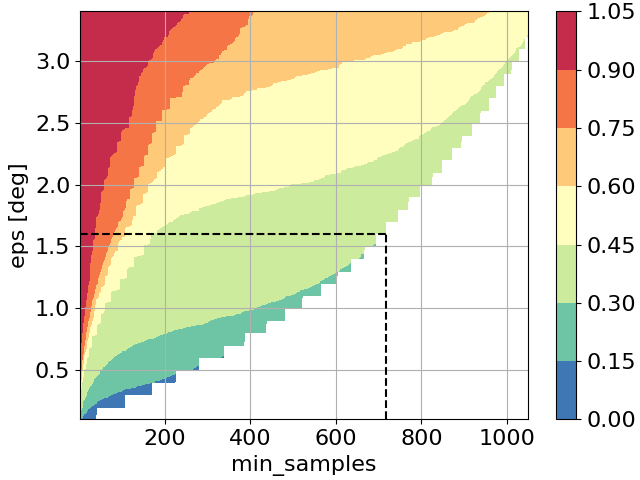}
\includegraphics[width=0.325\textwidth]{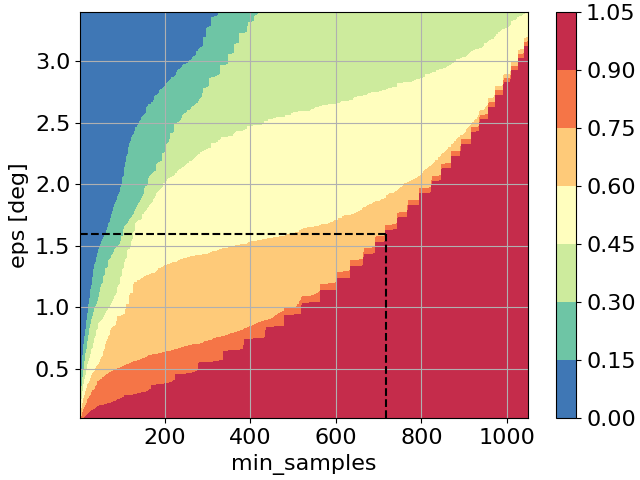}
\caption{\it{Results for \texttt{dbscan} clustering on the Coma dataset for a parameter space of $\emph{eps} \in \left[0.1, 3.5\right]^\circ$ in steps of $0.1^\circ$ and $\emph{min\_samples} \in \left[2, 1050 \right]$ in steps of 1 sample: number of clusters found (left), ratio of galaxies in the largest cluster to the total number of galaxies in the dataset (centre), ratio of galaxies in the background field to the total number of galaxies (right). The parameter combination with maximum curvature in the number of clusters (left) is marked in all three plots (black dashed lines): $\emph{eps}=(1.6\pm 0.1)^\circ$ and $\emph{min\_samples}=716 \pm 1$.}}
\label{fig:dbscan_scan}
\end{figure*}

Since we want to keep model assumptions at the necessary minimum, we run \texttt{dbscan} with all possible parameter combinations in $\emph{min\_samples} \in \left[ 2, 1050\right]$ in steps of 1 galaxy and $\emph{eps} \in \left[ 0.1,3.5\right]^\circ$ in steps of $0.1^\circ$. 
For each combination, we note the total number of clusters found in the dataset, the number of galaxies assigned to the largest cluster (as a ratio with respect to the total number of galaxies, 2477), and the number of galaxies not assigned to any cluster (again as a ratio). 
Fig.~\ref{fig:dbscan_scan} shows the results (from left to right). 
White regions in the left and centre plot denote parameter combinations that do not yield any clustering, such that all galaxies are assigned to the field (red colour in the right plot). 
Fig.~\ref{fig:dbscan_scan} (left) shows that, except for a very small parameter range, Coma is the only cluster detected by the algorithm. 
Altogether, the clustering hints at a small amount of substructures with lower number density than the Coma cluster itself is expected to have.
This result is also consistent with Fig.~\ref{fig:dbscan_scan} (centre) and (right), as the number of field galaxies increases when the amount of galaxies assigned to the cluster decreases.

\begin{figure}
\centering
\includegraphics[width=0.65\textwidth]{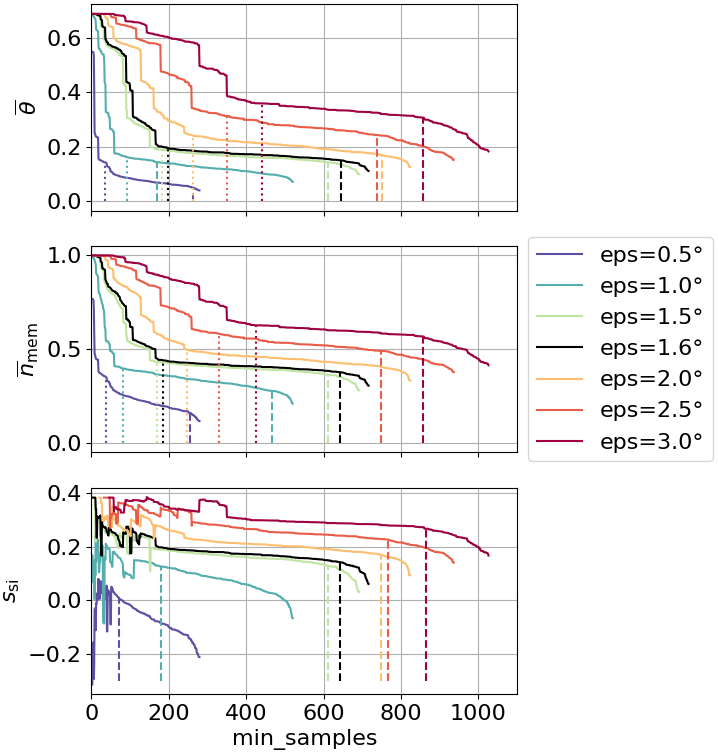}
\caption{\it{Quality of fit criteria for the parameter combinations shown in Fig.~\ref{fig:dbscan_scan}: normalised mean distance (top), normalised number of members in the largest cluster (centre), and the silhouette score (bottom) over \emph{min\_samples} for selected \emph{eps}-values. The \emph{min\_samples}-values with maximum curvature are also added {to mark the parameter ranges that robustly identify the cluster}: vertical dotted lines for least dense stable clustering, vertical dashed lines for most dense stable clustering. {Resulting clusterings for the parameter combinations marked by the dotted and dashed lines for \emph{eps}~$\in \left\{1.0, 1.5, 1.6, 2.0 \right\}$~degree are shown in Fig.~\ref{fig:dbscan_sky}.}}}
\label{fig:dbscan_qof}
\end{figure}

{To further constrain the optimum parameter combination, the algorithm \texttt{kneed}\footnote{https://github.com/arvkevi/kneed}, a Python package, \cite{bib:Satopaa2011}, is used to determine the point of maximum curvature from the border line between a single cluster and no clustering. Points of maximum curvature in general, also called "knees" or "elbows", mark the change in the clustering behaviour. Here, we determine the point at which the overdensity of Coma transits into the background density}, as will become clear in Fig.~\ref{fig:dbscan_qof}. It thus constrains a salient scale and characteristic density to identify the cluster.

This point is found at $\emph{eps}=(1.6\pm 0.1)^\circ$ and $\emph{min\_samples}=716 \pm 1$. 
The parameter combination is robustly recovered for the standard settings of \texttt{kneed} and also when varying the degree of polynomial data interpolation.
Yet, the point of maximum curvature depends on the parameter ranges over which the curve is considered. 
Varying the start and end points of the $\emph{eps}$ and $\emph{min\_samples}$ intervals, the position changes to slightly higher values when the interval size is decreased.

To rate the quality of the clustering and determine the most robust optimum parameter values, we investigate three possible quality criteria:
\begin{enumerate}
\item \textit{normalised mean distance}: we determine the mean distance between all points in the largest cluster which corresponds to the mean angular distance between the member galaxies on the sky. 
Then, it is normalised to the radius of the cone on the sky, $10^\circ$. 
This normalised mean distance is abbreviated as $\overline{\theta}$.
\item \textit{normalised number of cluster members}: we determine the number of members of the largest cluster and divide it by the total number of points in the dataset. This ratio is abbreviated as $\overline{n}_\mathrm{mem}$.
\item \textit{silhouette score}: we calculate this normalised summary statistic that is frequently used in signal processing, \cite{bib:Rousseeuw1987}. Its values range from -1 to 1, with 1 representing the optimum partitioning into clusters for the given dataset and -1 the worst possible. 
Each data point is assigned a score between -1 and 1 representing its fitness to be assigned to a certain cluster compared to being assigned to any other cluster. 
The average of all data points then results in the silhouette score of the entire dataset.
As the case considered here is a one-class clustering on top of background noise, the average is expected to be lower than for multi-class clustering without noise.
The silhouette score is abbreviated as $s_{\mathrm{si}}$.
\end{enumerate}

\begin{figure*}
\centering
\includegraphics[width=0.30\textwidth]{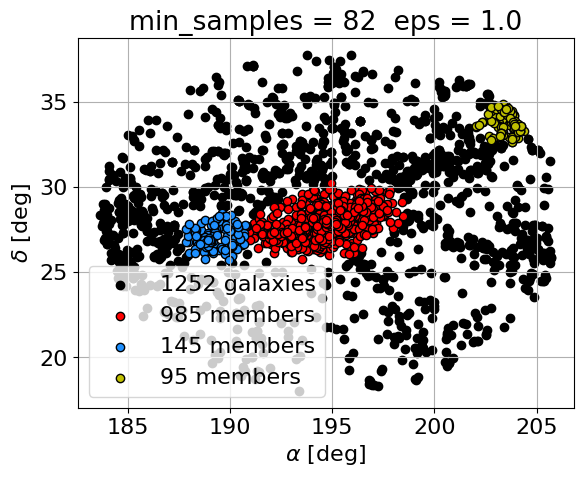}
\includegraphics[width=0.30\textwidth]{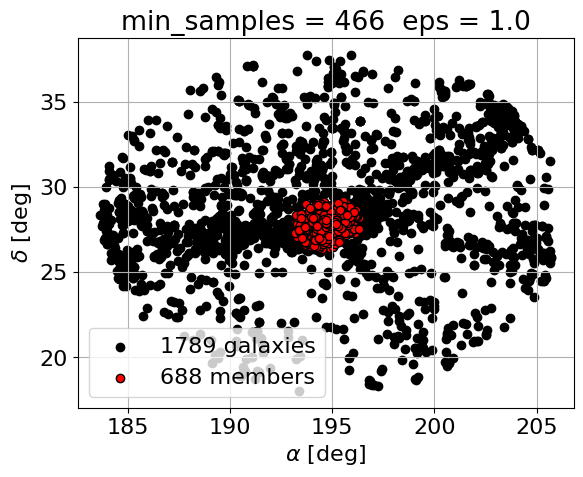}
\includegraphics[width=0.30\textwidth]{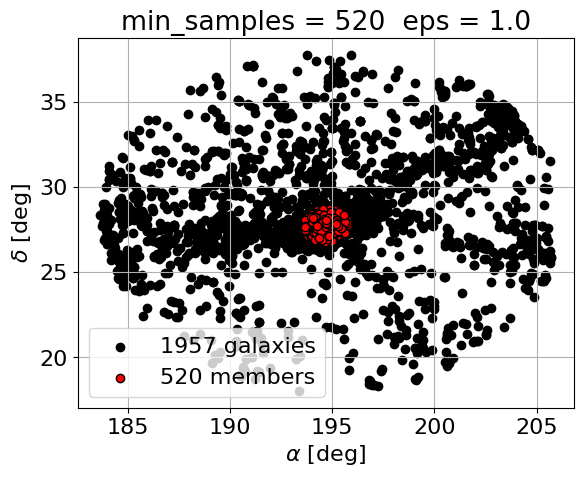}
\includegraphics[width=0.30\textwidth]{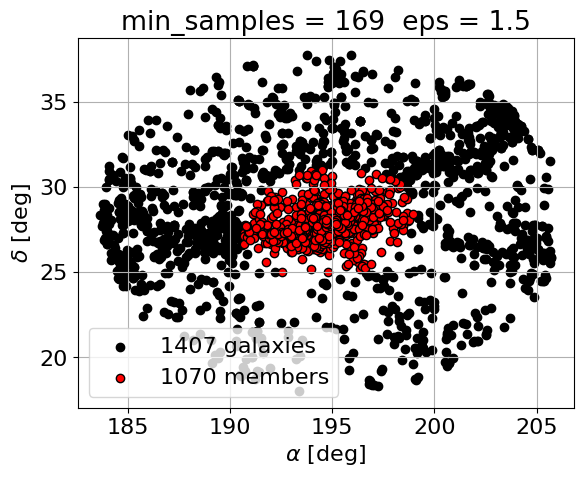}
\includegraphics[width=0.30\textwidth]{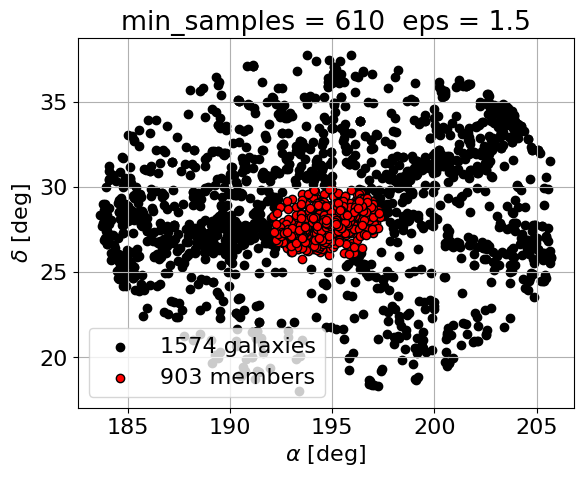}
\includegraphics[width=0.30\textwidth]{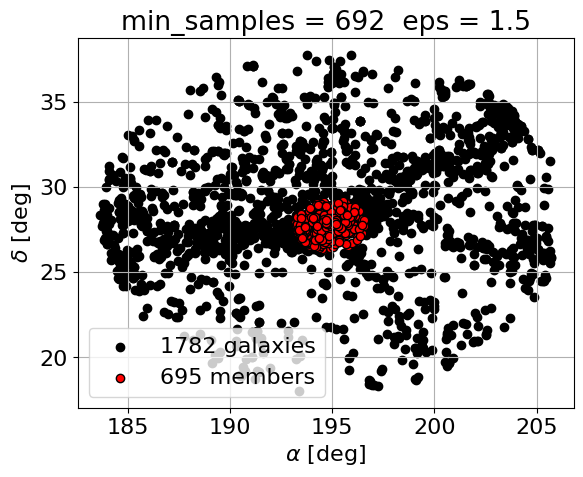}
\includegraphics[width=0.30\textwidth]{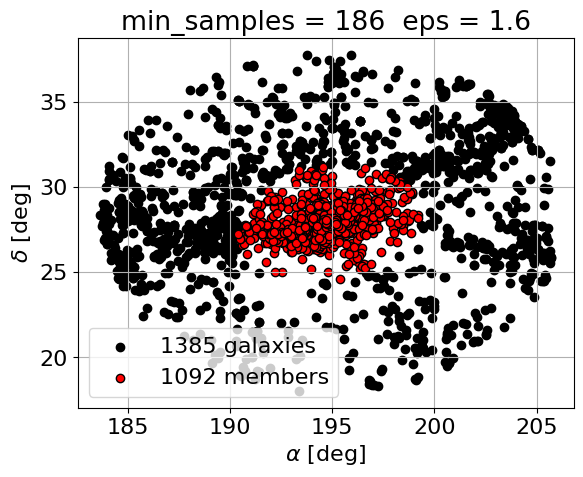}
\includegraphics[width=0.30\textwidth]{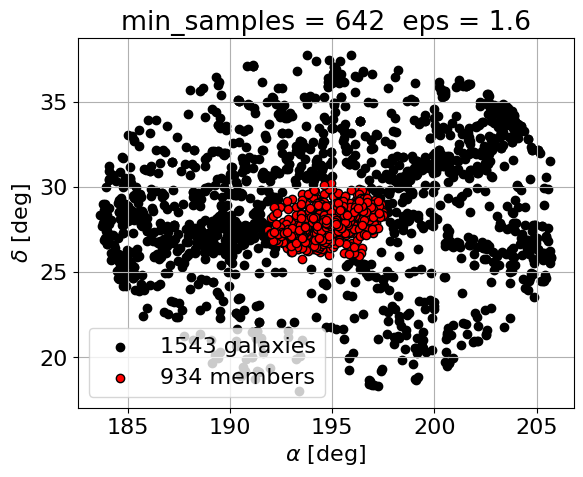}
\includegraphics[width=0.30\textwidth]{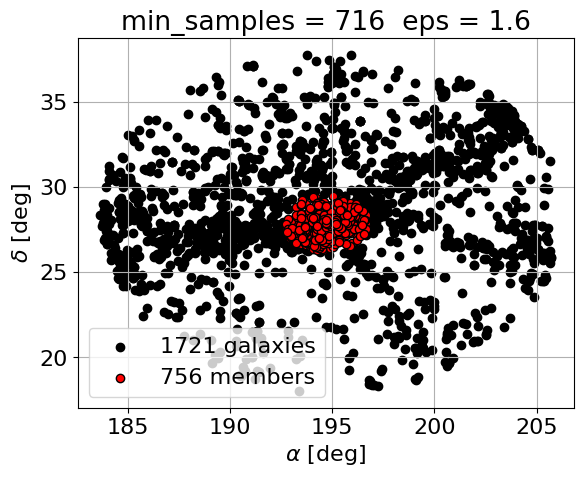}
\includegraphics[width=0.30\textwidth]{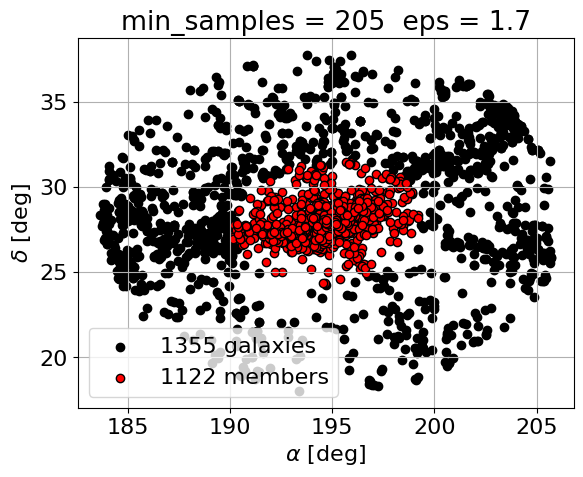}
\includegraphics[width=0.30\textwidth]{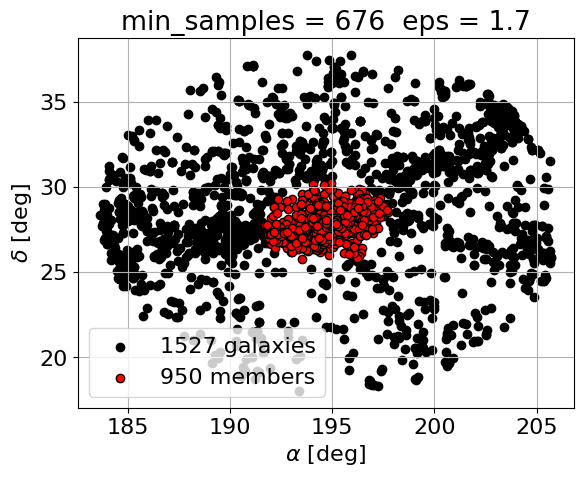}
\includegraphics[width=0.30\textwidth]{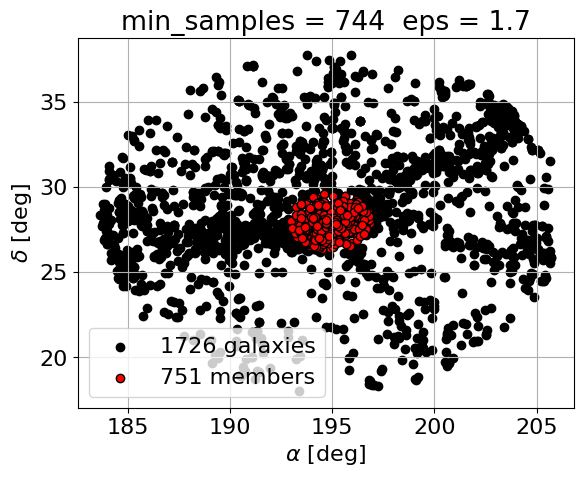}
\includegraphics[width=0.30\textwidth]{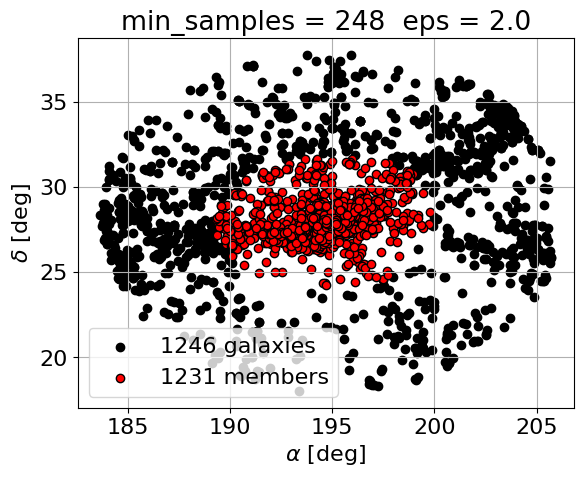}
\includegraphics[width=0.30\textwidth]{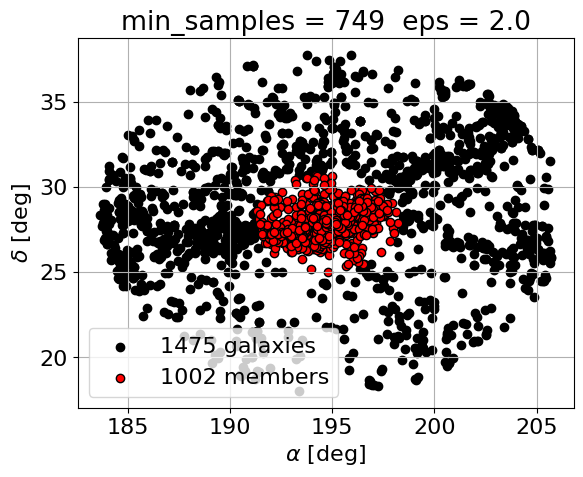}
\includegraphics[width=0.30\textwidth]{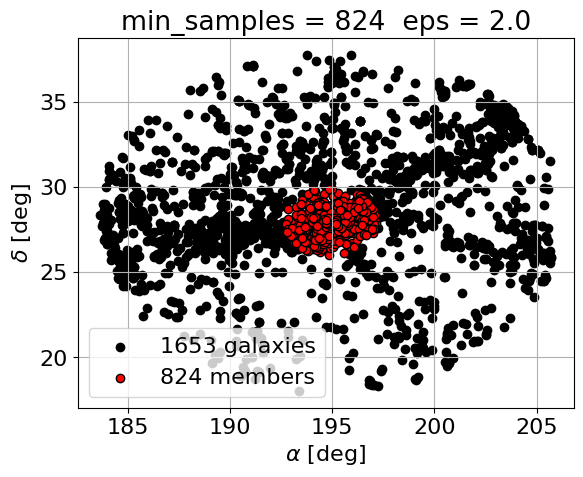}
\caption{\it{{Coma \texttt{dbscan} clustering results for different parameter-value combinations: least-dense robust clustering from $\overline{n}_\mathrm{mem}$ (left column), densest robust clustering from $\overline{n}_\mathrm{mem}$ (centre column), and maximum possible clustering (right column). The least-dense clustering may include structures accreted from the outskirts, while the maximum clustering under-estimates the number of members even for large \emph{eps}. The selected parameter combinations are marked by the vertical lines in Fig.~\ref{fig:dbscan_qof}.}}}
\label{fig:dbscan_sky}
\end{figure*}

{Fig.~\ref{fig:dbscan_qof} shows the three criteria plotted over all possible values of \emph{min\_samples} for selected \emph{eps}-values. 
It is apparent from the two top-most plots that increasing \emph{min\_samples} for fixed \emph{eps}-value leads to a denser clustering with smaller mean distance and a smaller number of cluster members. 
The steep decrease for small \emph{min\_samples} occurs because the algorithm mostly clusters the background density with many small-scale clusters inside but not identifying a stable, large structure. 
Only when the plateau is reached at intermediate \emph{min\_samples}, the large-scale cluster is robustly identified, which will also be shown in Fig.~\ref{fig:dbscan_sky} that shows the clustering results for selected parameter combinations}. 
Using \texttt{kneed} again in the default configuration, we determine the point of maximum curvature close to the first point on the plateau (as indicated by the dotted vertical lines in the two top-most plots in Fig.~\ref{fig:dbscan_qof}). 
We consider this parameter configuration for each \emph{eps}-value to be the least dense stable clustering. 
At high \emph{min\_samples}-values, we determine the last stable densest clustering by determining the point of maximum curvature in front of the last \emph{min\_samples}-value that still leads to a cluster being found (as indicated by the dashed vertical lines in the two top-most plots in Fig.~\ref{fig:dbscan_qof}). 
Since the plots of the silhouette score vary a lot at small \emph{min\_samples}-values, we can only determine the densest stable clustering from this quality criterion. 
A comparison of all three plots shows that $\overline{n}_\mathrm{mem}$ robustly identifies all points of extremum curvature and is thus the best criterion out of the three to select the optimum clustering parameters.
Fig.~\ref{fig:dbscan_sky} subsequently shows the on-sky clustering results for several \emph{eps}-values for the least-dense stable clustering, the densest stable clustering, and the maximum \emph{min\_samples}-value that yields a clustering as based on the $\overline{n}_\mathrm{mem}$ criterion. Hence, the extremum curvature points indeed identify salient parameter combinations that capture the extent of the cluster well.

We thus see from Figs.~\ref{fig:dbscan_qof} and \ref{fig:dbscan_sky} that the optimum parameter set is $\emph{eps}=1.6^\circ$ and $\emph{min\_samples}=642$, the densest robust clustering based on $\overline{n}_\mathrm{mem}$.
For lower or higher values of \emph{min\_samples} at fixed \emph{eps}, the algorithm seems to over- or under-estimate the extent of the cluster, respectively.
For lower \emph{eps}-values, the algorithm may not only under-estimate the number of members, but also cluster additional smaller-scale structures. 
The slope of the curve in Fig.~\ref{fig:dbscan_qof} falls over the entire range of \emph{min\_samples} without a clear plateau. 
Going to higher \emph{eps}-values, the plateau acquires a falling slope again and the maximum curvature at high \emph{min\_samples} is decreasing and the point of maximum curvature is hard to identify.

Thus, the final member selection is based on the optimum parameters found robustly and automatically by the algorithm as salient curvature extrema in Figs.~\ref{fig:dbscan_scan} and~\ref{fig:dbscan_qof}: we set $\emph{eps}=1.6^\circ$ as the size of the clustering-circle  {because this is the \emph{eps}-value for which the clustering behaviour changes in Fig.~\ref{fig:dbscan_scan} (left)}. 
For \emph{min\_samples}, we  {read off} three values  {from Fig.~\ref{fig:dbscan_qof} to partition the cluster into three parts}: the maximum $\emph{min\_samples}=716$  {for which the algorithm detects a cluster encompasses} all member galaxies in the core of the cluster, the densest robust clustering with $\emph{min\_samples}=642$ selects all members of the full cluster, and the least-dense robust clustering with $\emph{min\_samples}=169$ selects cluster member galaxies in the outskirts of the cluster. The latter region will be subject to further tests if these galaxies still belong to the cluster or are the transitioning region into the field. While other cluster studies are often interested in separating galaxy members by their type, blue or red galaxies, see, for instance, \cite{bib:Sohn2017}, we disregard this information and focus on the location of a member within the cluster. As will be detailed in Section~\ref{sec:final_selection}, the core region is the one with the highest number of galaxies that all Coma member sets have in common because most of them focus on the virialised region. The additional two outer regions are introduced to determine the zero-velocity surface of the cluster. Without adding further model assumptions about Coma's environment, this surface could be either located at the boundary of the full Coma set or at the outmost point of the outskirts. Which of those is more likely will only become clear when plotting the infall models in Section~\ref{sec:velocity_distance_relation} and comparing to Coma mass estimates from prior studies in Section~\ref{sec:mass_det}. 

The core of the cluster thus consists of 756 galaxies, the full cluster adds 178 galaxies, and the outskirts contain 158 galaxies, so that the total number of member galaxies amounts to 1092, as shown in Fig.~\ref{fig:dbscan_sky} (third row). The second and the fourth row in Fig.~\ref{fig:dbscan_sky} show the change in position and number of selected cluster members if this configuration is changed according to $\emph{eps}=(1.6\pm 0.1)^\circ$.

\subsection{Final Coma-member Selection}
\label{sec:final_selection}

\begin{figure*}
    \centering
\includegraphics[width=0.48\textwidth]{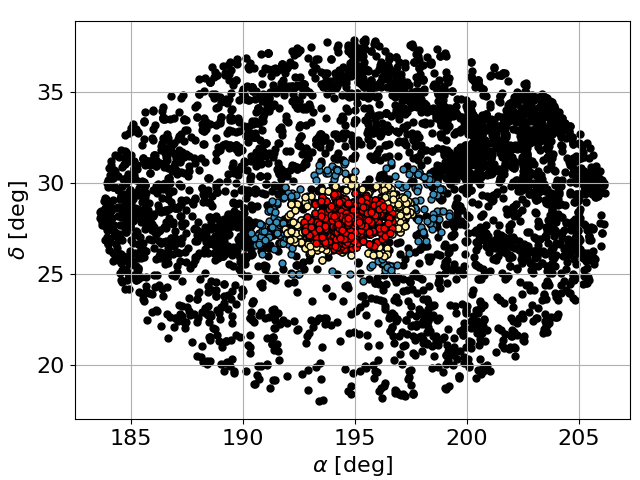} \hfill
\includegraphics[width=0.48\textwidth]{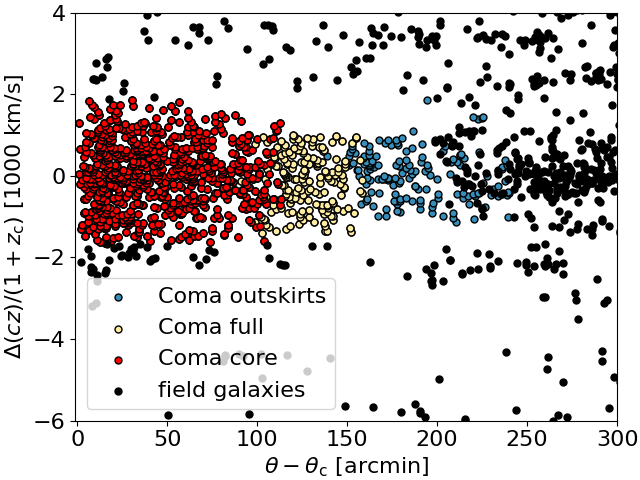}
\\
\includegraphics[width=0.48\textwidth]{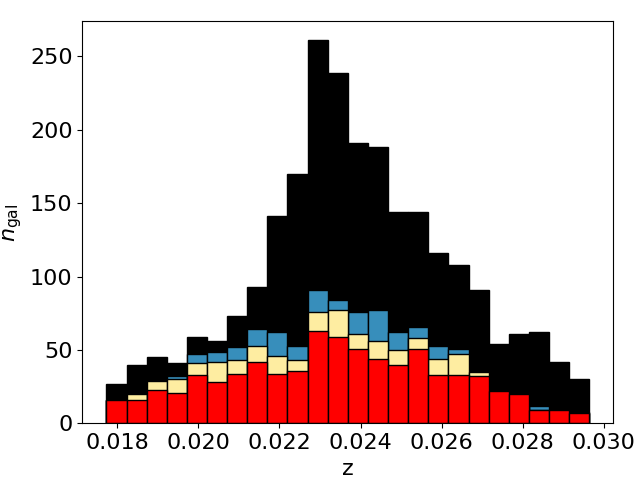} \hfill
\includegraphics[width=0.48\textwidth]{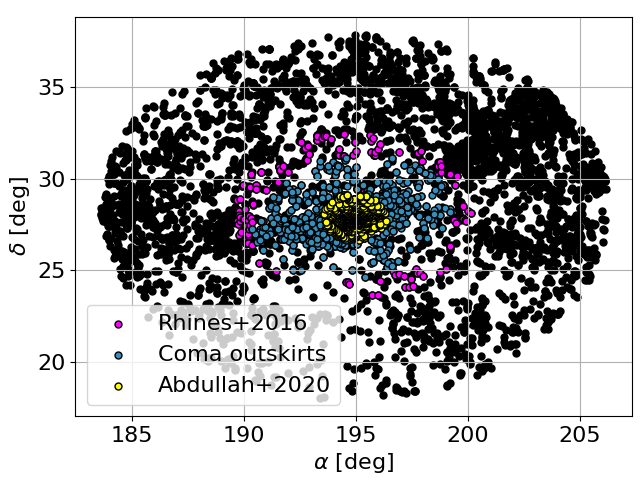}
\caption{\it{Final Coma member galaxy selection: sky distribution of member galaxies (top left), velocity-distance diagram for the angular distance on the sky from the cluster centre at $z_\mathrm{c}=0.0232$ (top right), redshift distribution (bottom left), and a comparison to two existing Coma member data sets from the literature (bottom right). Red coloured galaxies are in the core, yellow ones in the full dataset, blue ones in the outskirts of our Coma member selection. Field galaxies are in black.}}
\label{fig:mem_selection}
\end{figure*}

Combining all information for the selected Coma member galaxies from Sections~\ref{sec:selection_line_of_sight} and \ref{sec:selection_sky}, we arrive at the Coma dataset shown in Fig.~\ref{fig:mem_selection}. 
The top left plot shows the core, full, and outskirts members as identified by the \texttt{dbscan} clustering detailed in Section~\ref{sec:selection_sky} (red, yellow, and blue galaxies, respectively). With the same color scheme, we plot the distribution of the galaxies in a velocity-angular-distance diagram (top right).
As our member selection was performed for angular distances on the sky, the $x$-axis shows the radial distance from the centre in terms of these angular distances instead of physical ones.
The bottom left plot then shows a histogram of all galaxies in the redshift range of Coma.  

An update of the Coma cluster parameters based on our member selection, can be found in Tab.~\ref{tab:Coma_props} and a complete list of all Coma members is available in the online supplements to this paper. A comparison to values from the literature, added at the end of Tab.~\ref{tab:Coma_props}, shows overall agreement within the confidence bounds.
 {From Fig.~\ref{fig:mem_selection} (bottom right), we can also read off that our member selection has a large overlap with existing data sets and only differs at the boundaries. For instance, the Coma members according to \cite{bib:Abdullah2020} are completely absorbed in our Coma core region, while the Coma member selection of \cite{Rines_2016} extends to larger distances from the Coma centre than our dataset.}


\begin{sidewaystable}
\centering
\caption{{Synopsis of Coma cluster parameters. Based on our member selection (top three rows), $\Delta n_\mathrm{gal}$-values based on $\Delta \textit{eps} = \pm 0.1^\circ$ from Fig.~\ref{fig:dbscan_sky}, all $\alpha$ and $\delta$ values are given as the average of all selected member galaxies, uncertainties are 1-$\sigma$ standard deviations thereof. The $z_\mathrm{c}$- and $\mu$-values including their uncertainties are calculated as inverse-variance-weighted averages and their standard deviations (see Eq.~\eqref{eq:z_blue}). The $n_\mu$-column lists the number of galaxies in SDSS also having distance moduli $\mu$ from Cosmicsflows-4. Values from the literature (bottom rows) are based on the details in the given references. The final value used in this paper ("Outskirts") is marked. }}
\label{tab:Coma_props}
\vspace{8pt}

\begin{tabular}{|c|rr|ll|ll|ll|lll|}
\hline\hline
Data  & $n_\mathrm{gal}$ & $\Delta n_\mathrm{gal}$ & $z_\mathrm{c}$ & $\Delta z_\mathrm{c}$ & $\alpha_\mathrm{c}$ & $\Delta \alpha_\mathrm{c}$ & $\delta_\mathrm{c}$ & $\Delta \delta_\mathrm{c}$ & $n_\mu$ & $\mu_\mathrm{c}$ & $\Delta \mu_\mathrm{c}$ \\
 & & & & & [deg]  & [deg] & [deg] & [deg] & & [mag] & [mag] \\
\hline
``Core'' & 756 & $^{+0}_{-61}$ & 0.0234448 & 0.0000004 & 194.740407 & 0.782296 & 27.859534 & 0.578981 & 162 & 34.91 & 0.03 \\
``Full'' & 934 & $^{+16}_{-31}$ & 0.0232741 & 0.0000003 & 194.771145 & 1.134595 & 27.906406 & 0.727062 & 194 & 34.91 & 0.03 \\

\Xhline{1.2pt} 
\textbf{``Outskirts''} & \textbf{1092} & \textbf{$^{+30}_{-22}$} & \textbf{0.0233360} & \textbf{0.0000003} & \textbf{194.772150} & \textbf{1.553513} & \textbf{27.972992} & \textbf{0.952474} & \textbf{212} & \textbf{34.90} & \textbf{0.03} \\
\Xhline{1.2pt} 

SDSS \cite{bib:Abdullah2020} & 672 & & 0.0234 & & 194.93500 & &  27.912472 & & & & \\
DESI \cite{bib:Jimenez2024} & 2157 & & 0.0232 & & 194.95305 & & 27.980690 & & & & \\
DESI \cite{bib:Said2024} & 1696 & & 0.0231 & 0.02000 & 194.95292 & & 27.980555 & & & & \\
HeCS-SZ \cite{Rines_2016} & 1147 & & 0.0234 & & 194.92950  & & 27.938620 & & & & \\
SDSS \cite{bib:Sohn2017} & 1224 & & 0.0231 & & 194.92950  & & 27.938620 & & & & \\
\hline\hline
\end{tabular}
\end{sidewaystable}



\subsection{Cross-matching with Cosmicflows-4}
\label{sec:cross-matching}

\begin{figure*}
    \centering
\includegraphics[width=0.48\textwidth]{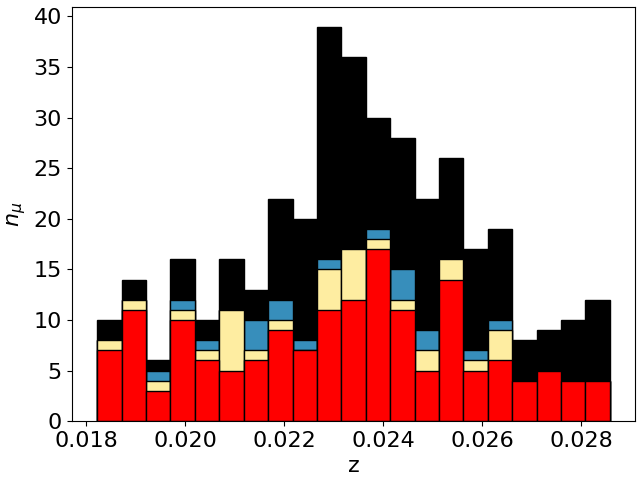} \hfill
\includegraphics[width=0.48\textwidth]{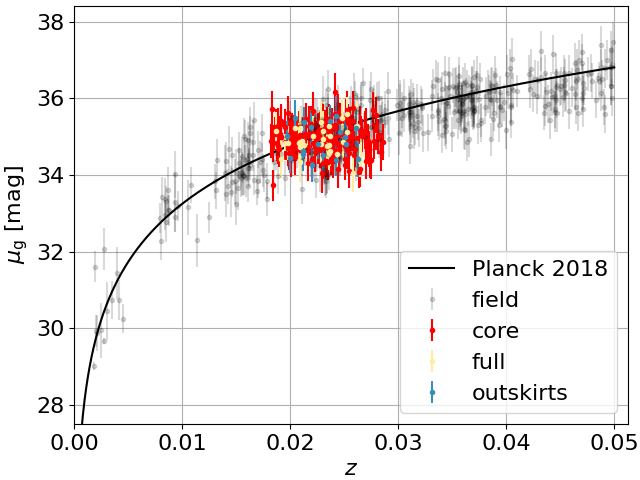}
\caption{\it{212 SDSS-selected Coma members and 479 field galaxies having CF4 distances: their SDSS redshift distribution (left), and CF4 distance moduli versus SDSS redshifts compared to an expected curve from the Planck 2018 cosmology (right). Red coloured galaxies are in the core, yellow ones in the full dataset, blue ones in the outskirts of our Coma member selection. Field galaxies are in black.}}
\label{fig:cf4_mems}
\end{figure*}

To be able to investigate the Hubble flow around Coma, we also need distances to the galaxies independent of the spectroscopic redshift measurements. Apart from DESI whose data is not fully publicly available, CF4~\cite{bib:Tully2023} is the most recent database that provides such distances. Therefore, we match our member selection and the field galaxies in the $10^\circ$ cone around Coma with the CF4 database\footnote{\url{https://edd.ifa.hawaii.edu}} to equip our SDSS member and field galaxies with independent distance estimates. Finding the best matches  {between the $\alpha$ and $\delta$ coordinates on the sky between SDSS and CF4} within one arcsec, distances for 212 {SDSS-selected} Coma members (162 from the core, 32 from the full, and 18 from the outskirts regions) and 479 field galaxies could be retrieved.     Fig.~\ref{fig:cf4_mems} summarises the results of the matching: the left plot shows the histogram of SDSS redshifts of the 212 member galaxies (coloured) and the 479 field galaxies (black) analogously to Fig.~\ref{fig:mem_selection} (bottom left), the right plot shows the CF4 distance moduli {with their observational uncertainties as error bars} versus the SDSS redshifts compared to an expected curve from the Planck 2018 cosmology.

\section{Velocity-distance relation}
\label{sec:velocity_distance_relation}

After the cosmology-independent data selection, we now proceed to set up the Hubble diagram. 
First, the centre of mass of the cluster needs to be determined and observed distance moduli and their uncertainties converted to physical distances. Subsequently, we elaborate on the biases due to the inhomogeneous distribution of the cluster members and the systematic uncertainties due to the joint calibration of different distance measures in CF4 {based on different astrophysical mechanisms} that are not taken into account in the statistical propagation of the observational uncertainties {detailed in Sections~\ref{sec:centre_of_mass} and \ref{sec:distances_cosmicflows}}.
Then, models for the three-dimensional infall velocities of the galaxies onto the cluster centre are developed.

\subsection{Centre-of-Mass Determination}
\label{sec:centre_of_mass}

The centre of mass of the cluster on the sky is determined as the average of the $\alpha$- and $\delta$-coordinates of the $n_\mathrm{gal} = 1092$ member galaxies (see also Table~\ref{tab:Coma_props}), assuming that the uncertainties in the celestial coordinates are negligible compared to other sources of errors. Implicitly, this averaging assumes that the galaxies as luminous tracers of the  {total} cluster mass sample the density profile well.
As already discussed in \cite{bib:Wagner2025}, if the total cluster mass is the sum of member galaxy masses only, treating all galaxies equally can introduce a bias. If the total cluster mass is dominated by an overall dark matter halo, however, the individual member galaxy masses are negligible and it is rather the member galaxy number and distribution within the dark matter halo that may create a bias. Varying the galaxy masses in $\left[10^{11},10^{13}\right]h^{-1}~M_\odot$, \cite{bib:Serra2011} found that the impact on the centre of mass for 100 clusters with a median mass of $10^{14}~h^{-1}~M_\odot$ is small compared to other sources of uncertainties, like projection effects. We therefore leave the model-dependent mass distribution for future work and focus on the impact of the projection effects.

As we can read off Table~\ref{tab:Coma_props}, $\alpha_\mathrm{c}$- and $\delta_\mathrm{c}$ agree within 1-$\sigma$ standard deviation over all three Coma member regions from the core to the outskirts. 
As expected for a sampling that increases less than the volume it covers, the standard deviation increases due to the thinning sampling from the core to the outskirts. 
The farthest distance on the sky that a point in the ``core'', ``full'', and ``outskirts'' set has from the centre of Coma as defined by all $n_\mathrm{gal}=1092$ galaxies is $1.89^\circ$, $2.64^\circ$, and $3.99^\circ$, respectively.
Hence, all member galaxies seem to fall within the small-angle approximation on the sky, implying that $\theta-\theta_\mathrm{c} \approx 0$. 

For structures obeying this small-angle approximation, the redshift of their centre can be determined as if all galaxies were lying along the same line of sight towards us as observers. Including the uncertainties of the redshift measurements, this amounts to {a central redshift $z_\mathrm{c}$ with its statistical uncertainty $\Delta z_\mathrm{c}$}
\begin{equation}
z_\mathrm{c} = \frac{\sum \limits_{j=1}^{n_\mathrm{gal}} \sigma_{z,j}^{-2} z_{j}}{\sum \limits_{j=1}^{n_\mathrm{gal}}\sigma_{z, j}^{-2}} \;, \quad \Delta z_\mathrm{c} = \left(\sum \limits_{j=1}^{n_\mathrm{gal}}\sigma_{z, j}^{-2}\right)^{-1/2} \;.
\label{eq:z_blue}
\end{equation}
Here, we assume that the given uncertainties on the SDSS redshifts $\sigma_\mathrm{z}$ are the standard deviations of a normal distribution and that the measurements are independent of each other, such that $z_\mathrm{c}$ can be calculated as the \emph{inverse-variance-weighted average} with its corresponding standard deviation, $\Delta z_\mathrm{c}$.
Under these conditions, Eq.~\eqref{eq:z_blue} yields the maximum likelihood estimate of the true value which is also the best linear unbiased estimator (BLUE). 
The $z_\mathrm{c}$-values for the Coma regions according to this formula are listed in Table~\ref{tab:Coma_props} and show that they all lie within the bounds of the first estimate $0.0232\pm0.00025$ {determined in Section~\ref{sec:selection_line_of_sight} with a small scatter and a tiny statistical} uncertainty which may be negligible compared to the potential bias due to the unknown relation between the member galaxy distribution and the total cluster mass distribution. 

{Dropping the approximation that all galaxies lie on a single line of sight, we can determine the centre-of-mass redshift analogously to \cite{bib:Wagner2025} from the redshifts of the galaxy ensemble weighted by their angular distance from the cluster centre on the sky, $\theta_{\mathrm{c},j}\equiv \theta_\mathrm{c} - \theta_j$:
\begin{eqnarray}
z_\mathrm{c,cos} = \frac{\sum \limits_{j=1}^{n_\mathrm{gal}} (\sigma_{z,j}\cos\theta_{\mathrm{c},j})^{-2} z_{j}\cos\theta_{\mathrm{c},j}}{\sum \limits_{j=1}^{n_\mathrm{gal}}(\sigma_{z, j}\cos\theta_{\mathrm{c},j})^{-2}} \;, \label{eq:z_cos_blue}
\quad \Delta z_\mathrm{c,cos} = \left(\sum \limits_{j=1}^{n_\mathrm{gal}}\left(\sigma_{z, j}\cos \theta_{\mathrm{c},j})\right)^{-2}\right)^{-1/2} \;.
\end{eqnarray}
These formulae yield $z_\mathrm{c,cos} = 0.0233234$ and its statistical uncertainty $\Delta z_\mathrm{c,cos}=3\times 10^{-7}$.
The difference to the redshift obtained by Eq.~\eqref{eq:z_blue} is of the order of $\Delta z \approx 10^{-5}$, such that both values do not agree within their statistical uncertainties. However, calculating the differences in the central redshift between the "core", "full", and "outskirts" clustering results, see Table~\ref{tab:Coma_props} for their values, $\Delta z \approx 10^{-4}$, such that the projection effects are one order of magnitude smaller than potential biases due to different cluster member selections.
Therefore, in the following, we adhere to the small-angle approximation, unless stated otherwise.
}

Converting the redshift of the Coma cluster centre into an accurate line-of-sight velocity is not trivial without an assumption about a global background cosmology or observations of the total matter distribution and its dynamics along Coma's line of sight, see, for instance \cite{bib:Heinesen2023}. 
As a first estimate, we assume that the Coma cluster centre is moving with the cosmic expansion, i.e.~there is no additional bulk flow. 
Then, the two most commonly used conversions from redshift into cosmic recession velocity are
\begin{eqnarray}
    v_\mathrm{c} &\approx& c\,z_\mathrm{c} \;, \\
    v_\mathrm{c} &\approx& \frac{cz_\mathrm{c}}{1+z_\mathrm{c}} \left( 1 +  \frac{1-q_0}{2} z_\mathrm{c} \right) \;,
\label{eq:vrecs}
\end{eqnarray}
as derived, for instance, in \cite{bib:Visser2004} and discussed in \cite{bib:Davis2014}.
As usual, $q_0$ is the cosmic deceleration parameter (see also \cite{bib:Dunajski2008} for details on the interpretation of $q_0$).
Comparing both approximations at the redshift of the Coma cluster, $z_\mathrm{c}=0.0233360$, the difference in $v_\mathrm{c}$ amounts to 36~km/s, which is of similar amplitude as the uncertainty due to the varying membership assignments.
As shown in \cite{bib:Davis2011}, this difference in the approximations is an order of magnitude smaller than the systematic uncertainty in the recession velocity of the cluster centre caused by a bulk flow due to the Local-Universe matter inhomogeneities resulting in locally preferred flow directions.
Hence, introducing a conservative bound of 4\% $v_\mathrm{c} \approx 280$ km/s to be attributed to such a bulk flow as the largest source of systematic uncertainty, the line-of-sight velocity of the centre of all $n_\mathrm{gal}=1092$ Coma member galaxies is then computed as
\begin{equation}
v_\mathrm{c} =  (6995.9 \pm {0.1_\mathrm{stat} \pm 279.8_\mathrm{sys}})~\mbox{km}/\mbox{s} \;.
\label{eq:v_c}
\end{equation} 
This result is supported by the more recent analyses in \cite{bib:Watkins2023} and \cite{bib:Watkins2025} as well, estimating a bulk flow with an amplitude of around 300~km/s for a distance of 100~Mpc away from us as observers.

\subsection{Distances from Cosmicflows-4}
\label{sec:distances_cosmicflows}

{We infer the $r_\mathrm{c}$ from the 212 CF4 distance moduli $\mu$ we have for Coma member galaxies.
Since this ensemble is a bit less than 20\% of all SDSS-selected member galaxies, we determine their centre on the sky as}
\begin{equation}
\alpha_\mathrm{c,CF4} = (194.81\pm 1.36)^{\circ} \;, \quad \delta_\mathrm{c, CF4} = (27.89 \pm 0.77)^{\circ}
\end{equation}
and their BLUE redshift according to Eq.~\eqref{eq:z_blue} as
\begin{equation}
z_\mathrm{c, CF4} = 0.0231450 \pm {0.0000006_\mathrm{stat}} \;.
\end{equation}
We note that the centre on the sky is shifted mostly in $\alpha$, but is still within the standard deviation of the values obtained for the entire SDSS-selected member galaxy set. 
Yet, the redshift estimate, based on the SDSS-redshifts of these 212 galaxies, which is lower than $v_\mathrm{c}$, yields a lower line-of-sight velocity compared to $v_\mathrm{c}$
\begin{equation}
v_\mathrm{c, CF4} = (6938.7 \pm {0.2_\mathrm{stat} \pm 277.5_\mathrm{sys}})~\mbox{km}/\mbox{s} \;,
\end{equation}
{which implies that} the subsample of member galaxies with CF4 distances is biased to lower redshifts compared to the full ensemble by 0.8\%{, yet is still within the confidence bounds set by the 4\% $v_\mathrm{c, CF4}$ systematic uncertainty of a local bulk flow}. 
Assuming that the $\mu$ are uncorrelated and the given uncertainties $\sigma_\mu$ are normally distributed, the BLUE based on these 212 Coma member galaxies can be calculated analogously as in Eq.~\eqref{eq:z_blue} to obtain $\mu_\mathrm{c}$ and $\Delta \mu_\mathrm{c}$ as listed in Table~\ref{tab:Coma_props}.
The value remains constant within its precision, if we restrict the calculation to the ``core'' or the ``full'' region. 
%
\begin{figure*}
    \centering
\includegraphics[width=0.42\textwidth]{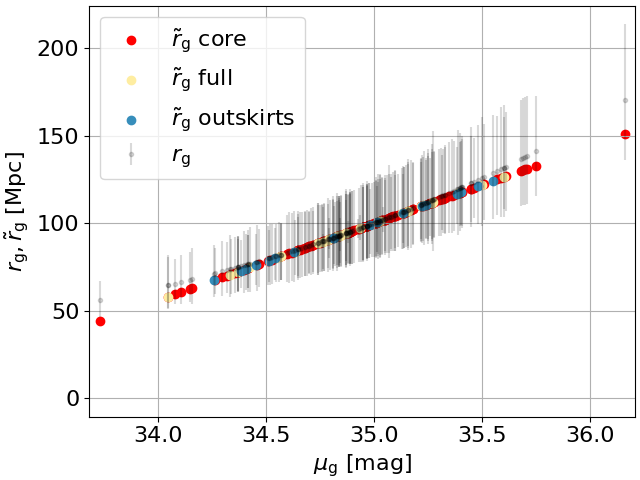} \hspace{4ex}
\includegraphics[width=0.42\textwidth]{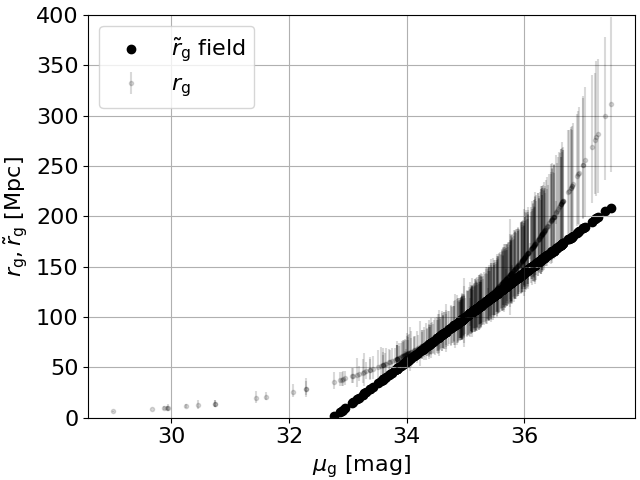}
\caption{\it{Testing the validity of Eq.~\eqref{eq:approx_dist} for the Coma member galaxies (left) and the field galaxies (right). $r_\mathrm{g}$ denotes the exact distance based on its Cosmicflows-4 $\mu_\mathrm{g}$ and error bars are drawn by propagating the $\sigma_\mu$ through to the distances, $\tilde{r}_\mathrm{g}$ denotes the approximation in Eq.~\eqref{eq:approx_dist}.}}
\label{fig:approx_dis}
\end{figure*}
We note that this $\mu_\mathrm{c, CF4}$ is obtained from the galaxies that altogether provided 2 surface brightness fluctuation (SBF) measurements, 6 supernova (SN) host distances, 26 Tully-Fisher (TF) relations, and 183 Fundamental Plane (FP) relations.
As further detailed in Section~10 of \cite{bib:Tully2023}, the $\mu$ from these different probes were calibrated with respect to their relative deviations, as well as their zero-points to minimise their spread across the entire CF4 volume and allow them to jointly probe the same cosmology. 
The zero-point of the joint dataset is lower than any of the individual probes.
As noted in \cite{bib:Tully2023} and supported by our findings here, selecting specific probes, like only FP or SN measurements, and choosing a local distance anchor, like Cepheids or Tip-of-the-Red-Giant-Branch (TRGB) observations, can introduce biases that are larger than the statistical uncertainties of the measurements. 

Converting $\mu_\mathrm{c}$ into $r_\mathrm{c}$ via $r = 10^{\mu_c/5-5}$, the non-linear relation causes the uncertainties in $r$ to become non-Gaussian. 
Hence, we can obtain $r_\mathrm{c}$ and its bounds by sampling from the normal distribution around $\mu_\mathrm{c}$ with standard deviation $\Delta \mu_\mathrm{c}$ and also compare this result to a propagation of uncertainties based on a truncated Taylor-series expansion
\begin{align}
r_\mathrm{c}= 95.63^{+1.23}_{-1.22} ~\mbox{Mpc} 
= 10^{\mu_\mathrm{c}/5-5} \pm \Delta \mu_\mathrm{c} \tfrac{\ln(10)}{5} 10^{\mu_\mathrm{c}/5-5}  = (95.70 \pm 1.20)~\mbox{Mpc}\;.
\label{eq:r_c}
%
\end{align}
Since both values are in good agreement with each other, the non-linearity is mild. 
This implies that the maximum difference between the distance moduli of the member galaxies to $\mu_\mathrm{c}$ is small, indeed $\left|\mu_i - \mu_\mathrm{c}\right|/\mu_\mathrm{c} \le 0.036$. The latter allows us to approximate the distance to each member galaxy~$j$ in terms of $r_\mathrm{c}$ and a correction as 
\begin{equation}
r_j = 10^{\tfrac{\mu_\mathrm{c}}{5} - 5 + \tfrac{\mu_j-\mu_\mathrm{c}}{5}} \approx r_\mathrm{c} \left(1 + \tfrac{\ln 10}{5} \left(\mu_j - \mu_\mathrm{c} \right) \right).
\label{eq:approx_dist}
\end{equation}
At the same time, we minimise the impact of the {large} measurement uncertainties on the distances and subsequently inferred quantities, as only relative distances between the cluster centre and each member galaxy occur in the formulae.

To investigate the size of the bias and check the validity of the approximation and its applicability to the 479 field galaxies around Coma having CF4 distance moduli, Fig~\ref{fig:approx_dis} shows the exact distances $r_\mathrm{g}$ including the slightly asymmetric error bars when propagating the 1-$\sigma$ uncertainties of $\mu_\mathrm{g}$ onto $r_\mathrm{g}$ (black lines) in comparison with the approximate distance (coloured dots for the Coma members in the left plot, black bold dots for the field galaxies in the right plot). As can be read off these plots, all Coma member galaxies but the one with the lowest magnitude support the approximation within their 1-$\sigma$ uncertainty. 
The farther the $\mu_\mathrm{g}$ from $\mu_\mathrm{c}$, the larger the bias to lower distances becomes. 
A similar, even more prominent trend is observed for the field galaxies.
The approximation only holds for 412 of the 479 field galaxies in the vicinity of $\mu_\mathrm{c}$. 
From Fig.~\ref{fig:approx_dis} (left), we read off that for a fixed distance modulus, the distance inferred from the approximation is biased low and, vice versa, for a fixed distance, the distance modulus of the approximation is biased high compared to the actual one. 
The approximation completely breaks down for distance moduli that are smaller than 33~mag, as the inferred distance goes to zero. 

The three-dimensional distance of each galaxy~$j$ to the centre of mass, as will be used in the galaxy infall models in Section~\ref{sec:radial_velocity}, is given as
\begin{equation}
r_{\mathrm{c},j}^2 = r_j^2 + r_\mathrm{c}^2 - 2 r_\mathrm{c} r_j \cos\theta_{\mathrm{c},j} \;.
\label{eq:rcj}
\end{equation}
For later use, we also introduce the normalised distances $\bar{r}_j \equiv r_j / r_{\mathrm{c},j}$ and $\bar{r}_\mathrm{c} \equiv r_\mathrm{c} / r_{\mathrm{c},j}$. Using the approximation of Eq.~\eqref{eq:approx_dist}, Eq.~\eqref{eq:rcj} can be approximated as
\begin{equation}
r_{\mathrm{c},j} \approx r_\mathrm{c} \, 2 \sin \left(\tfrac{\theta_{\mathrm{c}, j}}{2} \right) \left[1 + \tfrac{\ln 10}{5} \left(\mu_j - \mu_\mathrm{c} \right) \right] \;.
\end{equation}
Subsequently assuming $\theta_{\mathrm{c},j}$ is small, to leading order, $r_{\mathrm{c},j} \approx r_\mathrm{c} \cdot \theta_{\mathrm{c},j}$, so that the three-dimensional distance is approximated by the physical distance projected on the sky. This approximation resembles the decomposition in \cite{bib:Tsaprazi2025}, in which the distance moduli and redshifts consist of a cosmological part and one coming from peculiar velocities, inducing redshift space distortions for $z$. 


\subsection{Consistency of Calibrations}
\label{sec:consistency}
{We determine the local value of the Hubble constant by combining $v_\mathrm{c}$ from the SDSS-redshifts in Eq.~\eqref{eq:v_c} with $r_\mathrm{c}$ from the CF4 distances in Eq.~\eqref{eq:r_c} and we obtain}
\begin{equation}
H_0 = (73.10 \pm {0.92_\mathrm{stat} \pm 2.92_\mathrm{sys}})~\mbox{km}/\mbox{s}/\mbox{Mpc}\;.
\label{eq:H0}
\end{equation}
Comparing this small and purely statistical error with the systematic errors when assembling a single, consistent distance modulus measure for CF4, {which is up to 7~km/s/Mpc} (see Section~10 of \cite{bib:Tully2023} for details), we find that the dominant source of errors when joining all late-universe probes is systematics in the joint zero-point calibration and in the selection of absolute-distance anchors. 

\begin{figure}[t!]
    \centering
\includegraphics[width=0.42\textwidth]{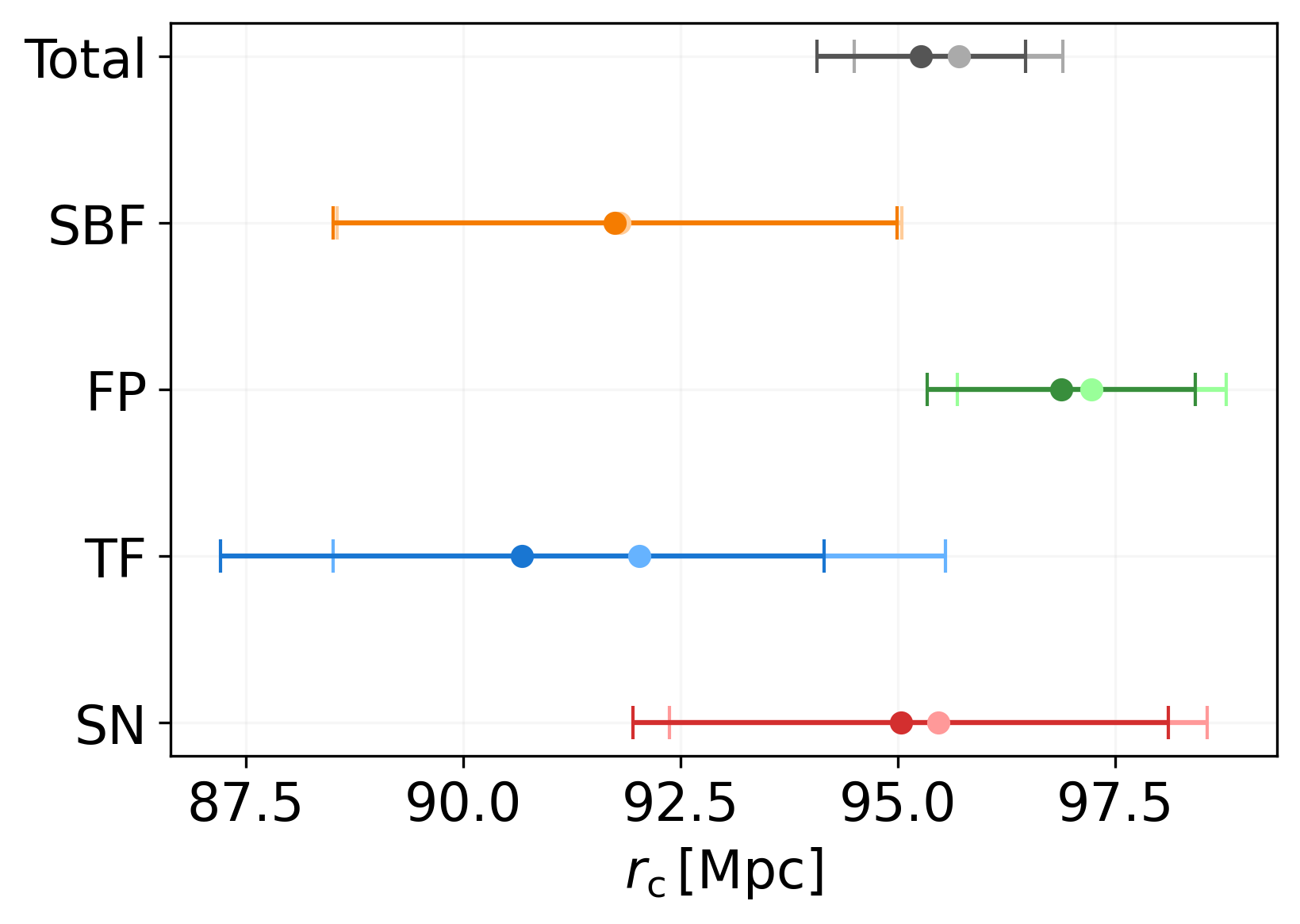}
\includegraphics[width=0.42\textwidth]{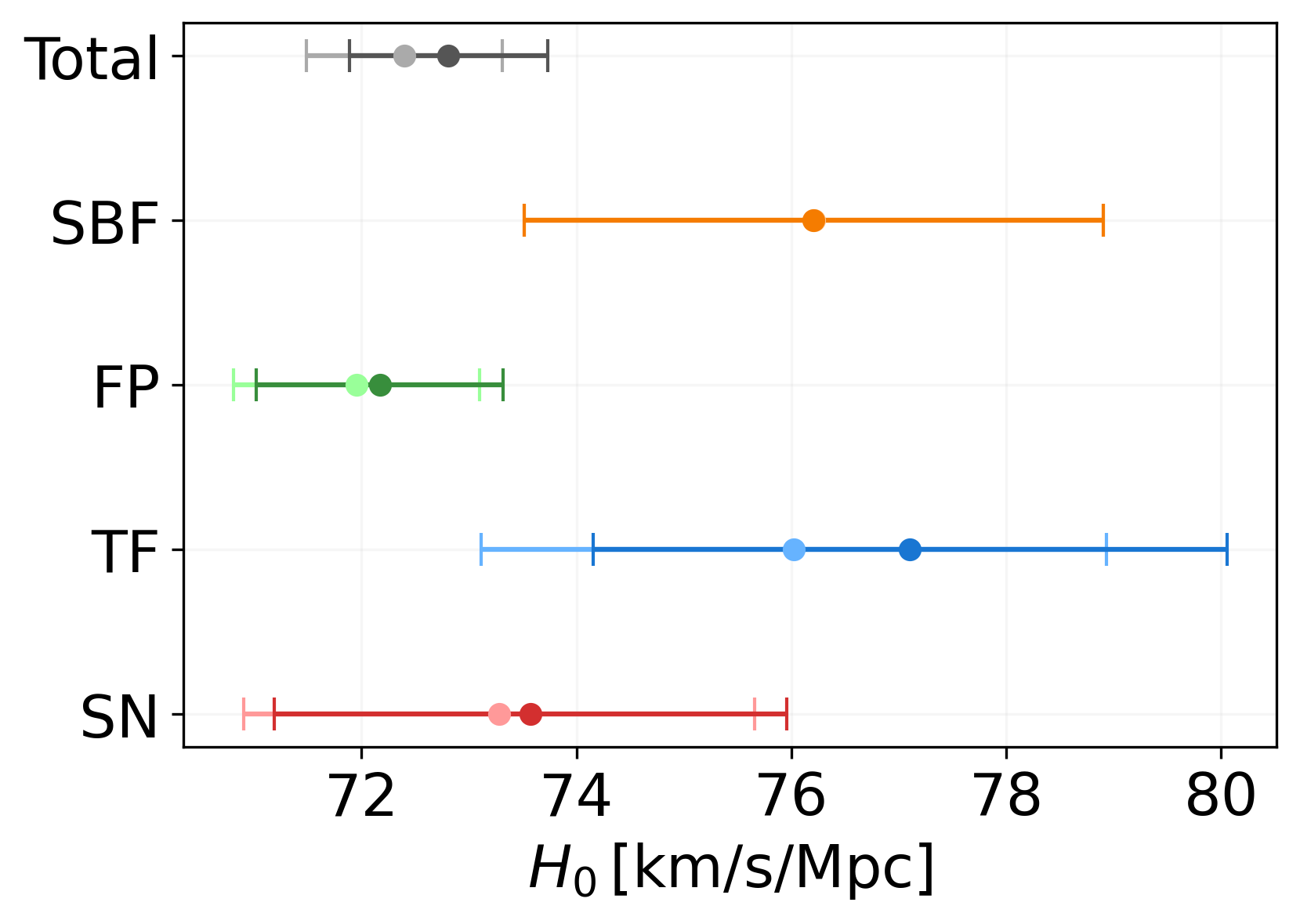}
\caption{\it{Distance to the Coma centre (left) from different probes: 2 Surface Brightness Fluctuations (SBF), 183 Fundamental Plane (FP), 26 Tully-Fisher (TF) and 6 Supernovae (SN) measurements in CF4. Pale colours are based on the small-angle approximation, dark colours account for the extend of Coma on the observer's sky. Hubble constant (right) inferred from these distances using Eq.~\eqref{eq:v_c} (pale colours) and {the velocity corresponding to Eq.~\eqref{eq:z_cos_blue}} (dark colours), respectively.}}
\label{fig:CF4_probes}
\end{figure}

Further investigating this issue, Fig.~\ref{fig:CF4_probes} shows $r_\mathrm{c}$ (top) and $H_0$ (bottom) calculated for the different physical distance probes. 
The ``Total'' values are obtained from all probes together, $r_\mathrm{c}$ according to Eq.~\eqref{eq:r_c} and the average {weighted for an extended structure instead of a single line of sight, analogous to} Eq.~\eqref{eq:z_cos_blue} in pale and dark colours, respectively.

The weighted average accounting for an extended structure on the sky always yields smaller $r_\mathrm{c}$-values than the average employing the small-angle approximation. 
Analogously, we calculate $r_\mathrm{c}$ for the 2 SBF, the 183 FP, 26 TF, and 6 SN distance measurements. 
All distance moduli measurements are based on the same distance anchors and zero-point-calibration that brings all CF4 data into consistency. 
This is why, for instance, $r_\mathrm{c}$ from the 6 SN is approximately $95.27$~Mpc by this calibration.
Yet, according to \cite{bib:Tully2023}, the SN in CF4 prefer a higher zero point than the total dataset, such that, shifting the distance moduli by 0.053 as reported in \cite{bib:Tully2023}, we obtain $98.07$~Mpc as average distance to the 6 SN. 
The latter value is well in agreement with the reported distance to 12 supernovae in Coma of $(98.5\pm 2.2)$~Mpc in \cite{bib:Scolnic2024} which were calibrated by the HST distance ladder.
A comparison of the redshifts and celestial coordinates shows that we share 3 supernovae.
Similarly, $r_\mathrm{c}$ from SBF measurements yields only a distance of $(91.8 \pm 3.2)$~Mpc, but the relative zero-point calibration according to \cite{bib:Tully2023} lowers their distance moduli even by 0.223 to bring SBF into consistency with all other probes in CF4. 
Reverting this relative calibration, the 2 SBF point at $r_\mathrm{c}=101.73$~Mpc.
Comparing the latter to the measurement by \cite{bib:Jensen2021}, yielding $(99.1\pm5.8)$~Mpc to NGC~4874 in Coma, our distance is in good agreement with this probe as well.
This is expected because NGC~4874 is one of the two SBF measurements in our dataset.

{We calculate the Hubble constants for the different probes, keeping $v_\mathrm{c}$ and $v_\mathrm{c,cos}$ fixed by Eqs.~\eqref{eq:v_c} and \eqref{eq:z_cos_blue} to investigate the impact of the small-angle approximation for the geometry, as well as the spread of $H_0$-values for the different probes. 
As for $r_\mathrm{c}$, we observe that the small-angle approximation only has a minor influence and the largest spread is caused by splitting the ensemble into the different probes.}
Even with a joint zero point calibration and distance anchor, the sparse statistics and inhomogeneous sampling of the cluster volume by the individual probes causes their average distances and average velocities (redshifts) to vary. 

Overall, our Coma member selection is consistent with state-of-the-art other works, using partly overlapping data as well. Yet, as shown, there is no percent-precision cosmological-model-independent distance to the centre of mass of Coma to be achieved because calibrating the different distance measurements based on different physical effects with respect to each other is still relying on too sparse a statistic, as also mentioned in \cite{bib:Tully2023}. 

\subsection{Radial Velocity and Hubble Flow}
\label{sec:radial_velocity}

\subsubsection{Radial Infall Models}

Next, we need to transform the observed velocities and positions to the centre of mass frame of Coma.  
Since the velocity components perpendicular to the line of sight are challenging to observe, the analysis is based on the observed line-of-sight components only. 
To calculate the radial infall velocities of the galaxies onto the Coma centre, assumptions about the system dynamics need to be added to estimate the impact of the unobserved velocity components, see \cite{bib:Karachentsev2010}. 
The impact of the two limiting models used to determine the radial infall velocities is further investigated in \cite{bib:Wagner2025}. 
For self-consistency, we briefly review the results here:
\begin{itemize}
\item The \textbf{minor infall model} approximates the radial infall velocity of a galaxy~$j$ onto the centre as  
\begin{align}
v_{\mathrm{r,min}} = v_{\mathrm{c}} \bar{r}_\mathrm{c} + v_{j}  \bar{r}_j - \cos \theta_{\mathrm{c},j}  \left( v_j  \bar{r}_\mathrm{c} + v_\mathrm{c} \bar{r}_j \right) \;.
\label{eq:v_min}
\end{align}
It thus treats the centre of mass and the galaxies symmetrically and assumes that $v_{\perp, \mathrm{c}} = v_{\perp,j} = 0$. 
(Instead, the perpendicular components could also conspire to a fine-tuned ratio, as detailed in \cite{bib:Wagner2025}, but this is the more unlikely scenario.)
\item The \textbf{major infall model} is an asymmetric model to approximate the radial infall velocity by projecting the velocity difference between a galaxy and the centre of mass onto the line of sight of the galaxy to obtain
\begin{align}
v_{\mathrm{r,maj}} = \frac{v_j - v_\mathrm{c} \cos \theta_{\mathrm{c},j}}{\overline{r}_j - \overline{r}_\mathrm{c} \cos \theta_{\mathrm{c},j}} \;. 
\label{eq:v_maj}
\end{align}
It thereby assumes that that $v_{\perp,\mathrm{c}} = v_{\mathrm{t}} = 0$, in which $v_\mathrm{t}$ denotes the tangential velocity. As for the minor infall model, there is a second, fine-tuned combination of these two unobservable velocity components, but it is more unlikely.
\end{itemize}
As already stated in Section~\ref{sec:centre_of_mass}, the models assume that the tracers have equal mass and are a representative sample of the underlying total mass distribution. To determine these infall models for the member galaxies of Coma, we can use the line-of-sight velocities as measured from SDSS spectra and $v_\mathrm{c}$ from Eq.~\eqref{eq:v_c}. We use the small-angle approximation $\theta_j-\theta_\mathrm{c}\approx 0$ for all galaxies~$j$ in the infall models.
For both cases, the leading-order approximation to the radial velocity is the difference of the measured velocity components along the line of sight, which amounts to a redshift difference times the speed of light for the small-angle approximation. 
The next order, quadratic in $\theta_{\mathrm{c},j}$, is then accounting for the geometry of the structure and therefore deviates for the two infall models. 

Fig.~\ref{fig:vec_dis_rel} shows the the velocity-distance relation from the centre of the Coma cluster using Eqs.~\eqref{eq:v_min} and \eqref{eq:v_maj} with error bars obtained from truncated Taylor expansions. 
As the distance is approximated with Eq.~\eqref{eq:approx_dist} and thus the physical distance projected on the sky, the plot resembles a stretched version of Fig.~\ref{fig:mem_selection} (top, right). 
The Coma member galaxies belonging to the ``core'', the ``full'', and the ``outskirts'' regions are sorted and not mixed as in Fig.~\ref{fig:cf4_mems} (right).
The field galaxies shown are selected such that their distance approximation of Eq.~\eqref{eq:approx_dist} is within the 1-$\sigma$ error bar of the distance inferred by their observed distance modulus (see Fig.~\ref{fig:approx_dis}, right). 

Comparing the velocity distributions between the minor and the major infall models, we note that the spread of the field galaxies is larger for the major infall model than for the minor, while for the cluster member galaxies, the difference is much smaller. 
This effect is caused by the ratio of velocities over distances in the major infall model, which amplifies a small spread in velocity differences more than the minor infall model. 

\subsubsection{Hubble Flow}

\begin{figure}
    \centering
\includegraphics[width=0.48\textwidth]{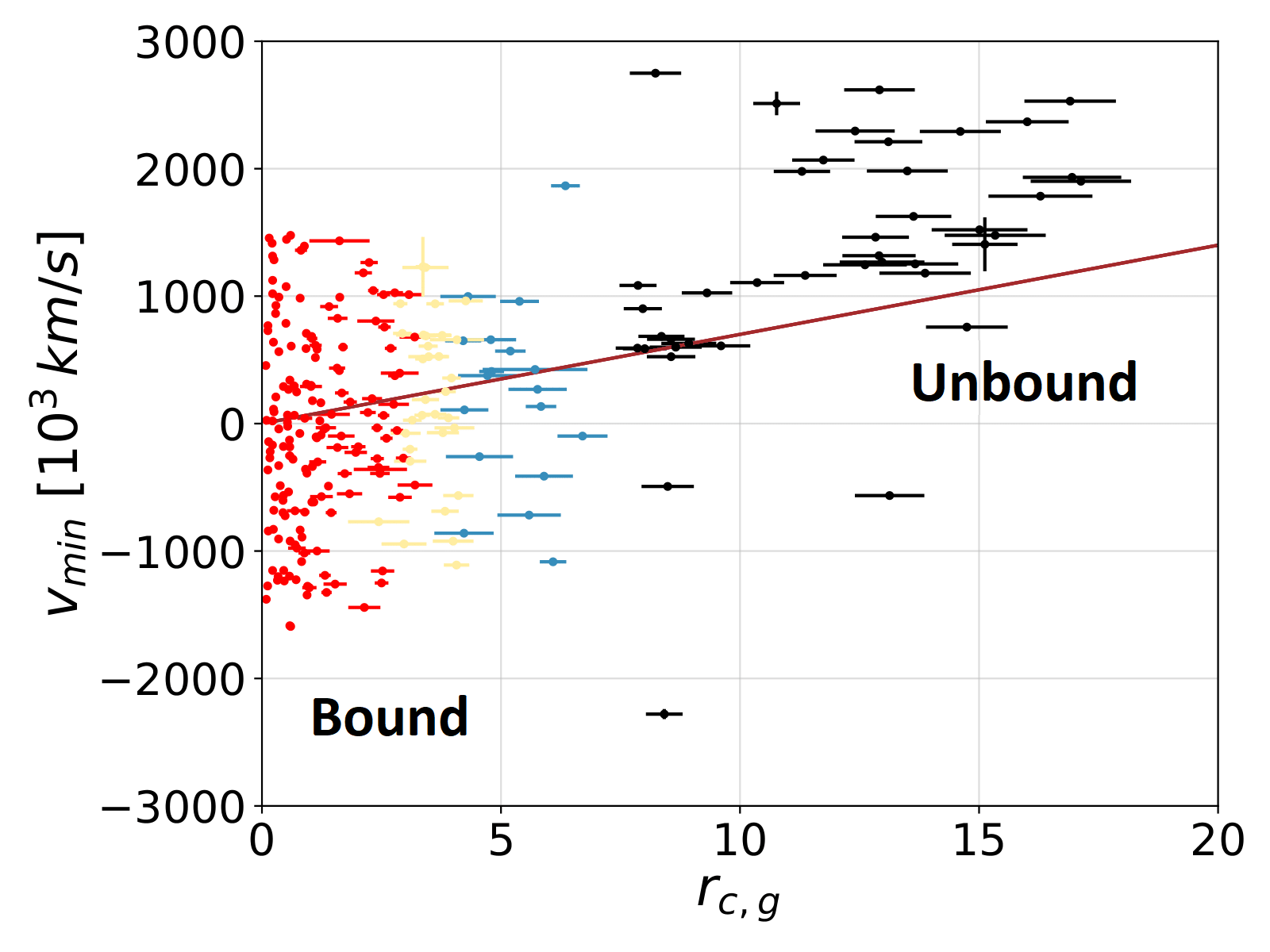}
\includegraphics[width=0.48\textwidth]{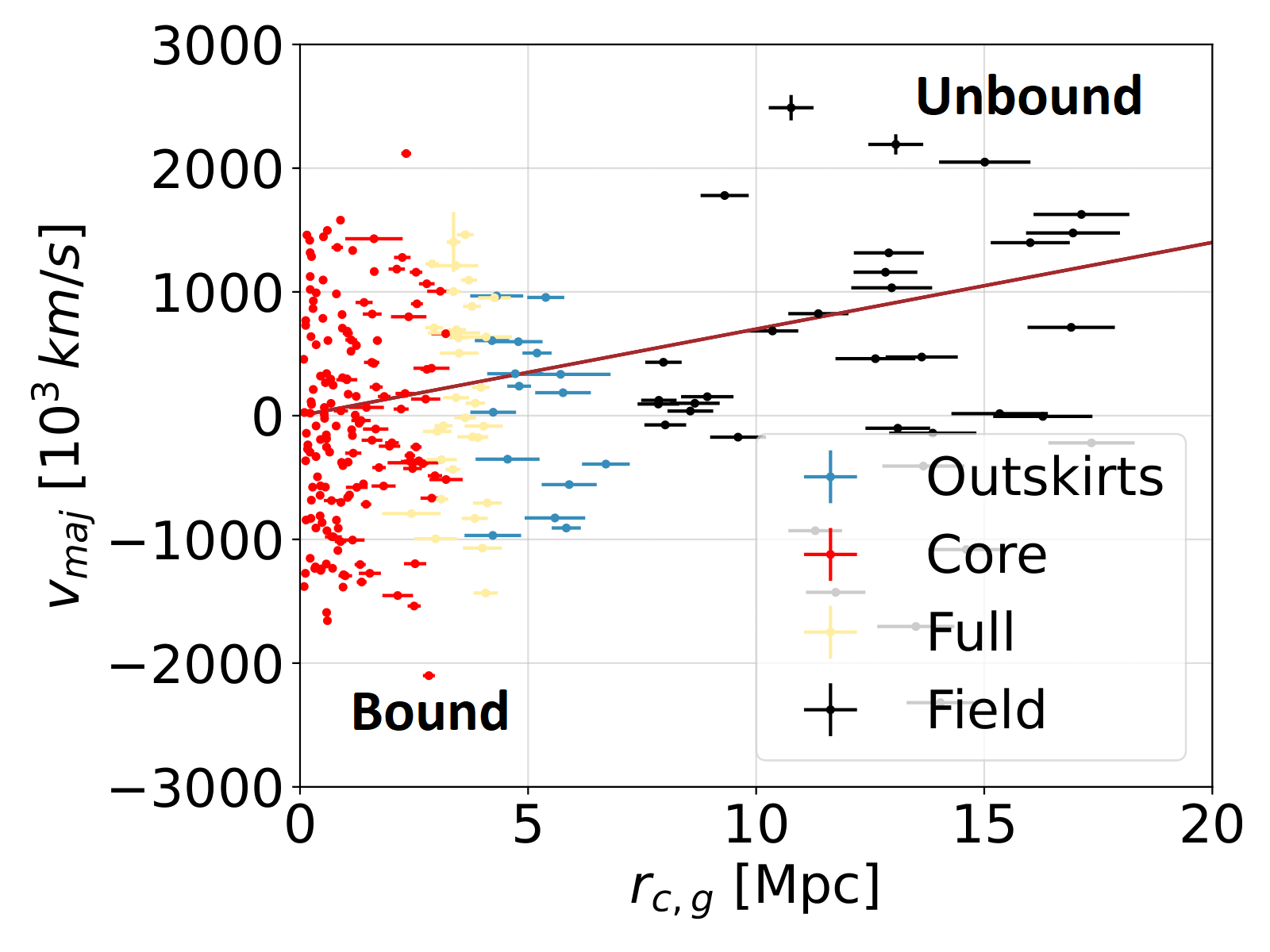} 
\caption{\it{
{Velocity-distance relation from the Coma centre for the minor (left) and the major infall model (right). The distance is the physical distance via Eq.~\eqref{eq:approx_dist}, infall velocities are obtained from Eqs.~\eqref{eq:v_min} and \eqref{eq:v_maj}. The bounded galaxies are marked in different colors, while the galaxies in the unbounded area are in black. 
}}}
\label{fig:vec_dis_rel}
\end{figure}

{The Hubble flow describes the motion of galaxies due to the expansion of the Universe close to cosmic structures which collapsed and decoupled from this expansion under their own gravitational attraction. \cite{bib:Sandage1986} introduced a method to estimate the mass of a system by its decelerating impact on the cosmic expansion, meaning its Hubble flow. 
Leaving $H_0$ as another free parameter, it becomes possible to jointly constrain the mass of the bound structure and $H_0$ from an observed velocity-distance diagram.
Although both quantities are degenerate with each other, self-consistent solutions for a mass from gravitationally-bound member objects and an $H_0$ from a far-field fit to objects in the Hubble flow can be determined. 
This has been done for galaxy groups, for instance, in \cite{bib:Peirani2006,bib:Peirani2008,bib:Penarrubia2014,bib:Sorce2016,Teerikorpi:2010} and \cite{bib:Kim2020, bib:Nasonova2011,bib:DelPopolo2022}. }


{Fig.~\ref{fig:vec_dis_rel} shows the Hubble flow for $H_0 = 70$~km/s/Mpc. As the plot shows, the core is part from the bounded area while the galaxies outside the core are on the Hubble flow around the Coma structure. Moreover, the quality cuts made above only allow us to identify 19 galaxies in the Hubble flow around Coma, which is additionally cluttered by infalling structures and the embedding into a filament. The Hubble flow to the Coma cluster region is not well-constrained and a cold Hubble flow cannot be properly identified because Coma is not isolated enough. Moreover the core of Coma is not concetrated around one point, that gives a hint for the dark matter density of the core, that is determined in the next section.}

\section{Mass Determination}
\label{sec:mass_det}

\subsection{Hubble Flow enclosed mass}
For a simple model of a point mass surrounded by test particles, \cite{bib:Sandage1986, bib:Karachentsev2009} correlates the mass to the turnaround (also called zero-velocity surface), via the relation  
\begin{equation}
M_{\mathrm{ta}} = \frac{\pi^2 r_{\mathrm{ta}}^3}{8 G t_U^2} \;,
\label{eq:mass_hubble}
\end{equation}
with the turnaround radius $r_\mathrm{ta}$ and the inverse age of the Universe $t_U$ being a function of the matter parameter $\Omega_\mathrm{m}$.
Inserting $\Omega_\mathrm{m}=0.3111$ from \cite{bib:Planck2018}, we obtain 
\begin{equation}
M_\mathrm{ta} = 4.42 \times 10^{12} \left( \frac{r_\mathrm{ta}}{1~\mbox{Mpc}} \right)^3 h^2 \, M_\odot
\end{equation}
as the total enclosed mass around the bounded area, \cite{Korkidis:2019nzk}. Since the transition region between gravitationally bound member galaxies and the Hubble flow is sampled with about 100 galaxies, we can estimate the turnaround radius for Coma based on our member selection set up in Section~\ref{sec:selection_sky}.
Since our clustering algorithm detected a gap in the number density of galaxies at $3.99^\circ$ on the sky from the ``outskirts'' of Coma to the field around the cluster, we estimate the projected approximation to the turnaround radius, Eq.~\eqref{eq:approx_dist}, from this angle on the sky via Eq.~\eqref{eq:mass_hubble} to be
\begin{equation}
r_{\text{ta,min}} \approx 4.87 \, h^{-1}~\mbox{Mpc} \;,
\label{eq:rta_min}
\end{equation}
which is about $6.66~\mbox{Mpc}$ using $H_0$ from \cite{bib:Tully2023}. This value gives a lower bound on the turnaround radius if all points in the ``outskirts" actually belong to Coma because the actual turnaround radius could also be located in the gap between the ``outskirts" and the field galaxies in the Hubble flow. The minimum distance of the field galaxies in the Hubble flow is 
\begin{equation}
r_{\text{ta,max}} \approx 8.63 \, h^{-1}~\mbox{Mpc} \;,
\label{eq:rta_max}
\end{equation}
{which is about $11.82~\mbox{Mpc}$ using $H_0$ from~\cite{bib:Tully2023}. This is an upper bound on $r_\mathrm{ta}$. Inserting the upper and lower bounds on the turnaround radius, Eqs.~\eqref{eq:rta_min} and \eqref{eq:rta_max}, the enclosed estimated mass yields} 
\begin{align} 
M_{\mathrm{ta}} &\in \left[5.11, 28.44\right] \times 10^{14}\,h^{-1}\,M_{\odot} \\ 
&= \left[6.99, 38.91 \right] \times 10^{14} \,M_{\odot} \;
\end{align}
We determine more accurate mass estimates from the caustic method and the virial theorem in the following.

\subsection{Caustics}
\label{sec:caustics}

The caustic technique is independent of the dynamical state of the cluster. 
It quantifies the combined effect of the velocity phase space, represented by a ratio $\gbeta$ involving the velocity anisotropy profile, and its gravitational potential, $\Phi(r)$, through the so-called caustic amplitude $\CauA$, which tracks the maximum observable escape velocity, $v_{\rm{esc}}^{2}(r)$, \cite{Diaferio:1997mq, Gifford:2013ufa}. 
This approach provides a robust method for estimating the mass of galaxy clusters without assuming a density profile, such as a Navarro-Frenk-White (NFW) profile, \cite{Navarro:1995iw}. 
Yet, it also requires a careful selection of members, namely those close to the escape-velocity boundary. 
Hence, interlopers need to be discarded, usually done according to \cite{Hartog1996}, and galaxies close to the cluster centre with a small infall velocity will not contribute much to the caustic surface. 

\begin{figure}[ht]
    \centering
    \includegraphics[scale=0.45]{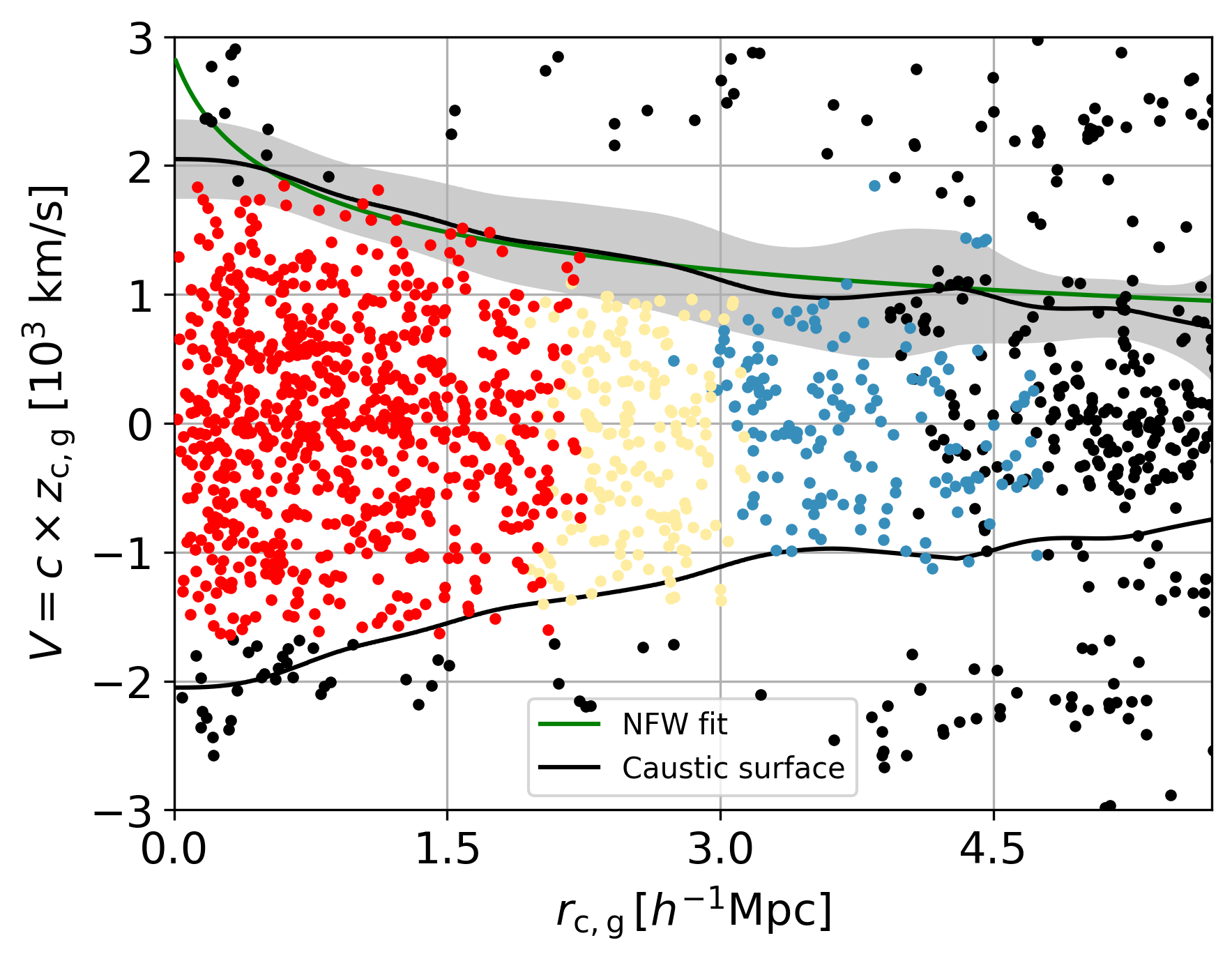}
    \\[-0.5ex]
    \includegraphics[scale=0.45]{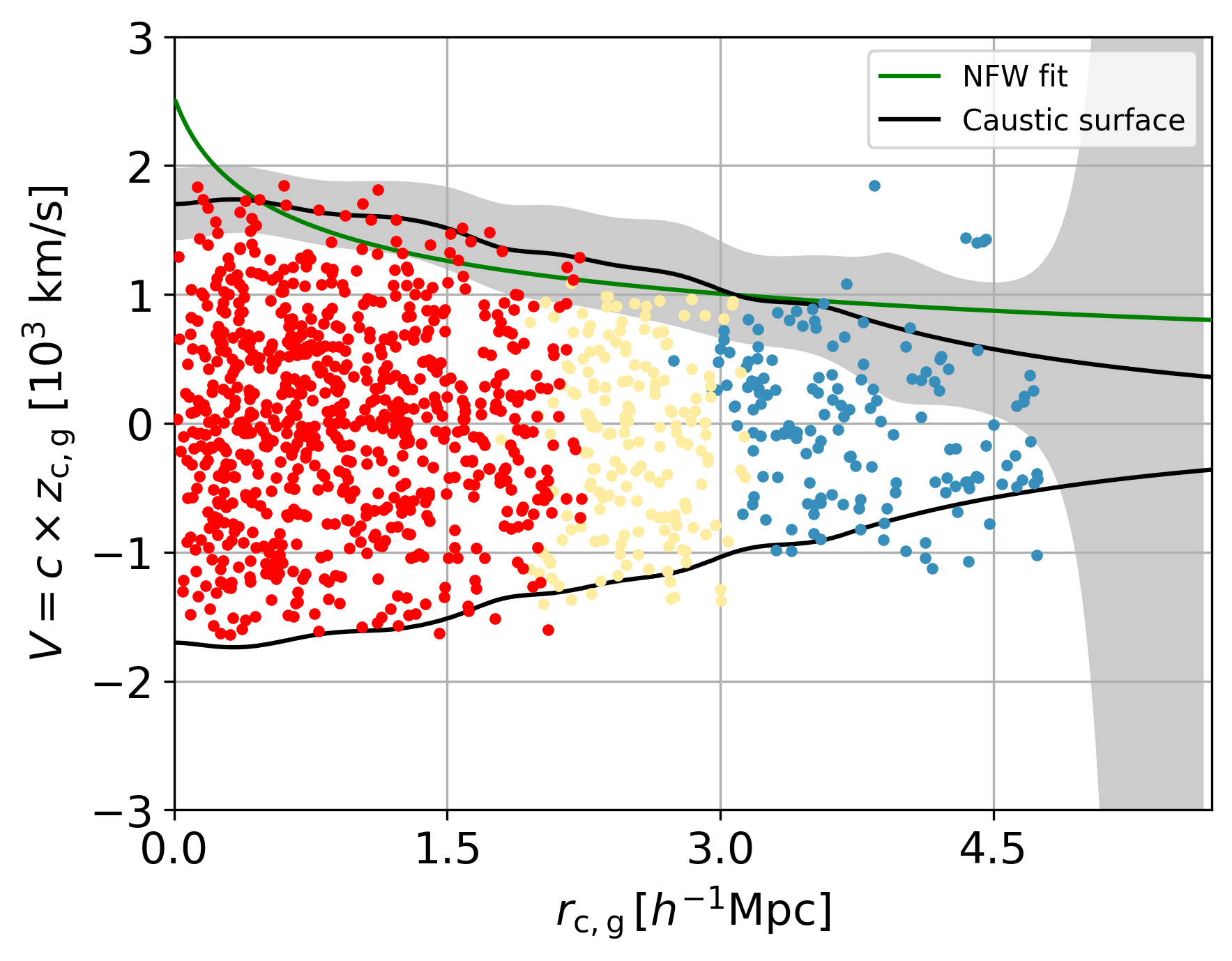}
    \\[-0.5ex]
    \includegraphics[scale=0.45]{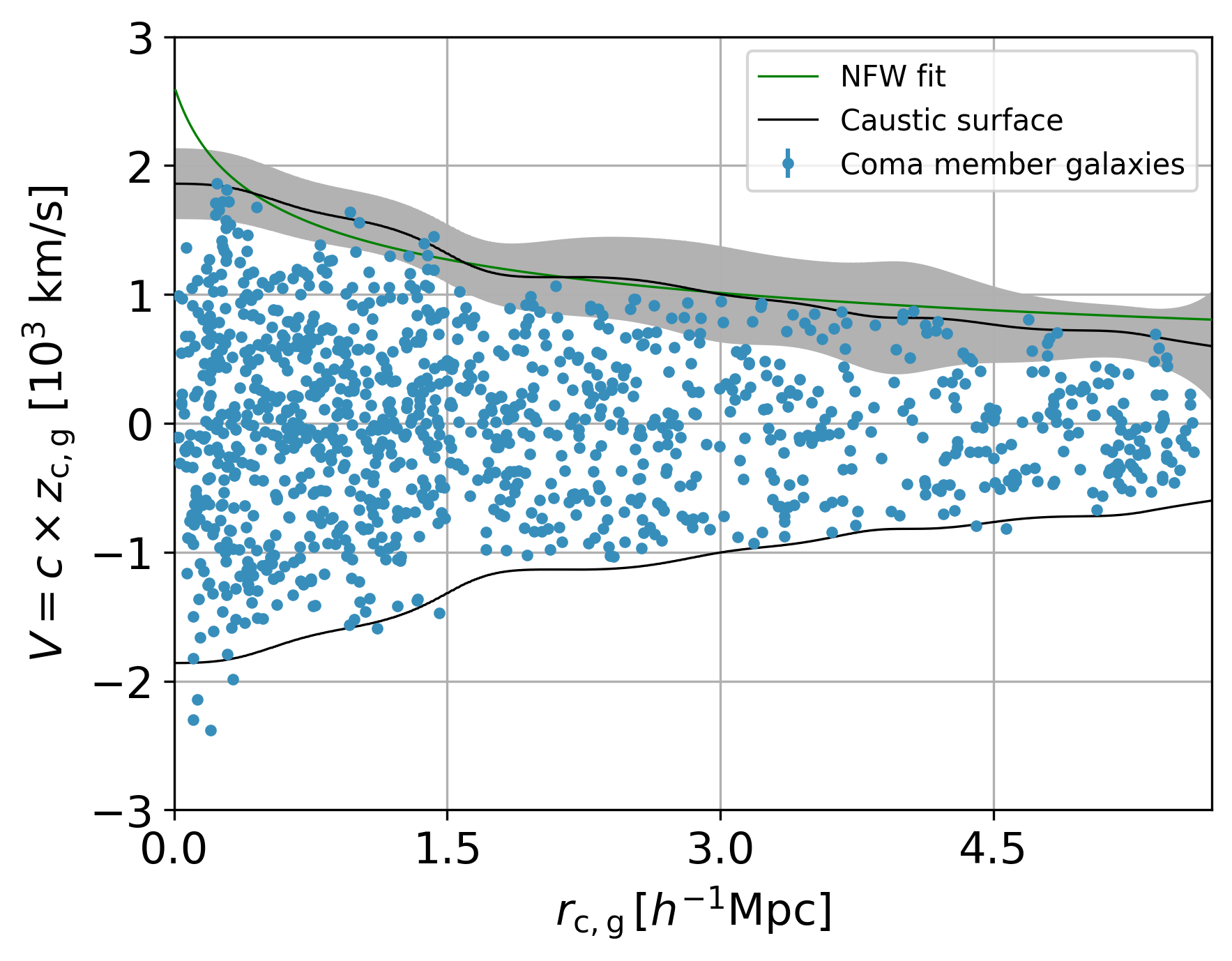}
    \vspace{-1.5ex}
     \caption{\it{Caustic surface (black line) for the 1092 Coma members selected in Section~\ref{sec:member_selection} with corresponding 1-$\sigma$ bounds (grey-shaded regions) and NFW-fit to the caustic surface (green line), including (top) and not including the field galaxies (centre), similarly, for 1147 Coma members as selected by \cite{Rines_2016} for comparison (bottom). Member galaxies are plotted in one colour, as \cite{Rines_2016} do not distinguish regions within Coma.}}
    \label{fig:caustic_surface_full}
\end{figure}

The gravitational potential of a profile for a bound object can be related to the caustic amplitude as 
\begin{eqnarray*}
    -2 \Phi(r) = \CauA \gbeta \;, \quad
\gbeta = \frac{3 - 2 \beta(r)}{1 - \beta(r)} \;,
\end{eqnarray*}
with the velocity anisotropy profile of the member galaxies $\beta(r) \equiv 1- \langle v_{\theta}^2 +v_{\phi}^2\rangle/2\langle v_{r}^2\rangle$. 
The total enclosed mass of a spherical system can then be estimated as
\begin{equation}
        \label{eqn:Mass_Profile}
         G M(<r)=\int_{0}^{r}\mathcal{A}^{2}(r)\mathcal{F}_{\beta}(r) dr \;,
\end{equation}
with $\mathcal{F}_{\beta}(r) \equiv \mathcal{F}(r) g_{\beta}(r)$ and $\mathcal{F}(r) \equiv -2\pi G {\rho(r)r^{2}}/{\Phi(r)}$. 
The filling factor $\mathcal{F}_{\beta}(r)$, is assumed to be a constant in the outer regions of the cluster, \cite{Diaferio:1997mq}. 
Common values vary between $0.5\leq \mathcal{F}_{\beta} \leq 0.7$ in the literature, see, for instance, \citep{Diaferio_2005, Diaferio09, Gifford:2016plw}. 
In the current analysis we assume $\mathcal{F}_{\beta} = 0.5$\footnote{See \cite{Butt:2024jes,Butt:2024civ} for a discussion on the estimation of the filling factor and the incurred mass bias with other probes such as hydrostatics. The value of $\mathcal{F}_{\beta}(r)$ can range between $\sim 0.8$ around the $r_{200}$ to $\sim 0.5$ in the outer regions extending beyond the virial range, which leads to reduced mass bias and allows a joint assessment of different probes.} to estimate the caustic mass\footnote{For this purpose, we utilize a modified version of the code presented in \cite{Gifford:2013ufa}.}, for ease of comparison with the estimates in literature.
For comparison, we also fit a NFW profile to the $v(r)$- and $M(r)$-relations given by our data. 

As complementary validation of our Coma member selection of Section~\ref{sec:member_selection}, the caustic technique itself can be utilized as a proxy for the selection of member galaxies by estimating the galaxies which lie within the caustic surface which we show in the top panel of Fig.~\ref{fig:caustic_surface_full}. In \cite{bib:Serra2013}, it has been demonstrated that the caustic surface is capable of accurately selecting the cluster members up to $95 \%$ extending out to the infall regions ($\sim r_{200}$).
Employing this method, we estimate the number of galaxy members to be $n_{\rm gal} = 1246$, which is comparable to the number of members assigned to Coma by both member selection approaches, ours and the one by \cite{Rines_2016} that we also analyse for comparison (see also Table~\ref{tab:mass_cons}).
The caustic method provides about $\sim 12\%$ more members. Fig.~\ref{fig:caustic_surface_full} (top) shows the resulting caustics and an NFW-fit to the caustic surface of our dataset. 

Subsequently, Fig.~\ref{fig:caustic_surface_full} (centre) shows the caustic surface estimated for all of our galaxy members $n_{\rm gal} = 1092$ (see Tab.~\ref{tab:Coma_props}) without including any field galaxies as in the top plot.
As only 212 member galaxies have independent cosmic distance measurements, we convert the $\theta_{\mathrm{c},j}$ into approximate physical distances by $\theta_{\mathrm{c},j} r_\mathrm{c}$.
The plot shows that the caustic surface is tightly constrained for the ``full Coma'' dataset. At larger distances, the error in the determination of the caustic surface becomes too large as the dataset is truncated in comparison to the inclusion of the field galaxies at larger radii. 
These plots can be compared to the one obtained for the Coma member selection detailed in \cite{Rines_2016}, shown in Fig.~\ref{fig:caustic_surface_full} (bottom).
A good overall agreement on the shape of the caustic is obtained, which is not surprising, given that the member selection has an overlap of 91\% (997 member galaxies).
From the surplus number of members identified by \cite{Rines_2016}, 132 are field galaxies in our selection, as they are located in the outskirts of our ``Coma outskirts''. 
Only 18 galaxies of \cite{Rines_2016} do not have a counterpart in our Coma light cone and thus must originate from a different source than SDSS~DR17. 

A comparison of Fig.~\ref{fig:caustic_surface_full} (centre) and (bottom) reveals that our member selection algorithm with a minimum amount of astrophysical assumptions about the dataset is similarly good at removing field galaxies and interlopers from the cluster as the ones from \cite{Rines_2016} employing the interloper removal according to \cite{Hartog1996}, see also \cite{Gifford:2013ufa, bib:Sohn2017, bib:Serra2013, bib:mamon2010, Wojtak:2006aq}. 
Comparing the NFW-profile fits of the two member selections, we note that the caustics inferred by \cite{Rines_2016} yield a better match to the NFW profile compared to ours. 

\begin{figure}
\centering
\includegraphics[scale=0.5]{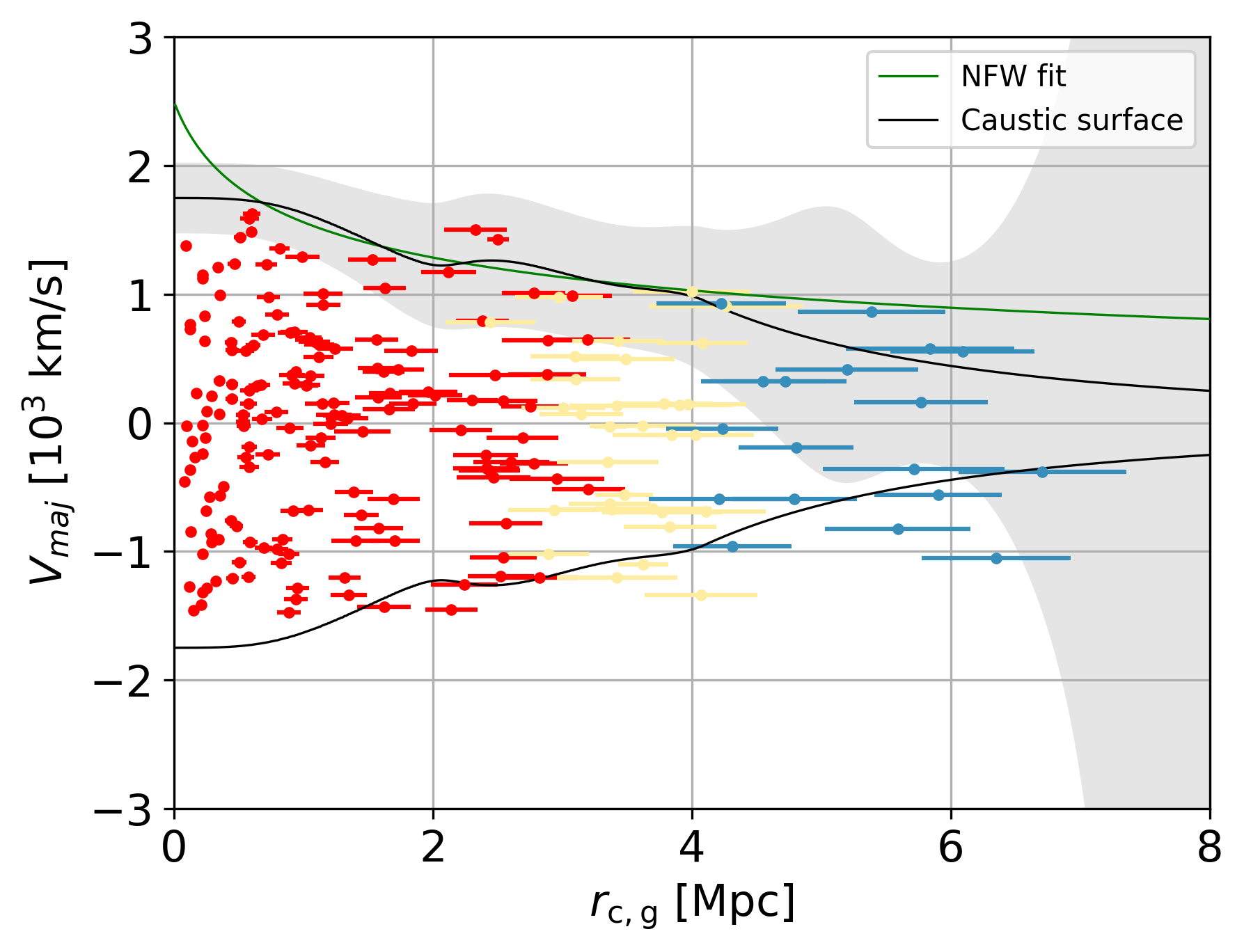}
\includegraphics[scale=0.5]{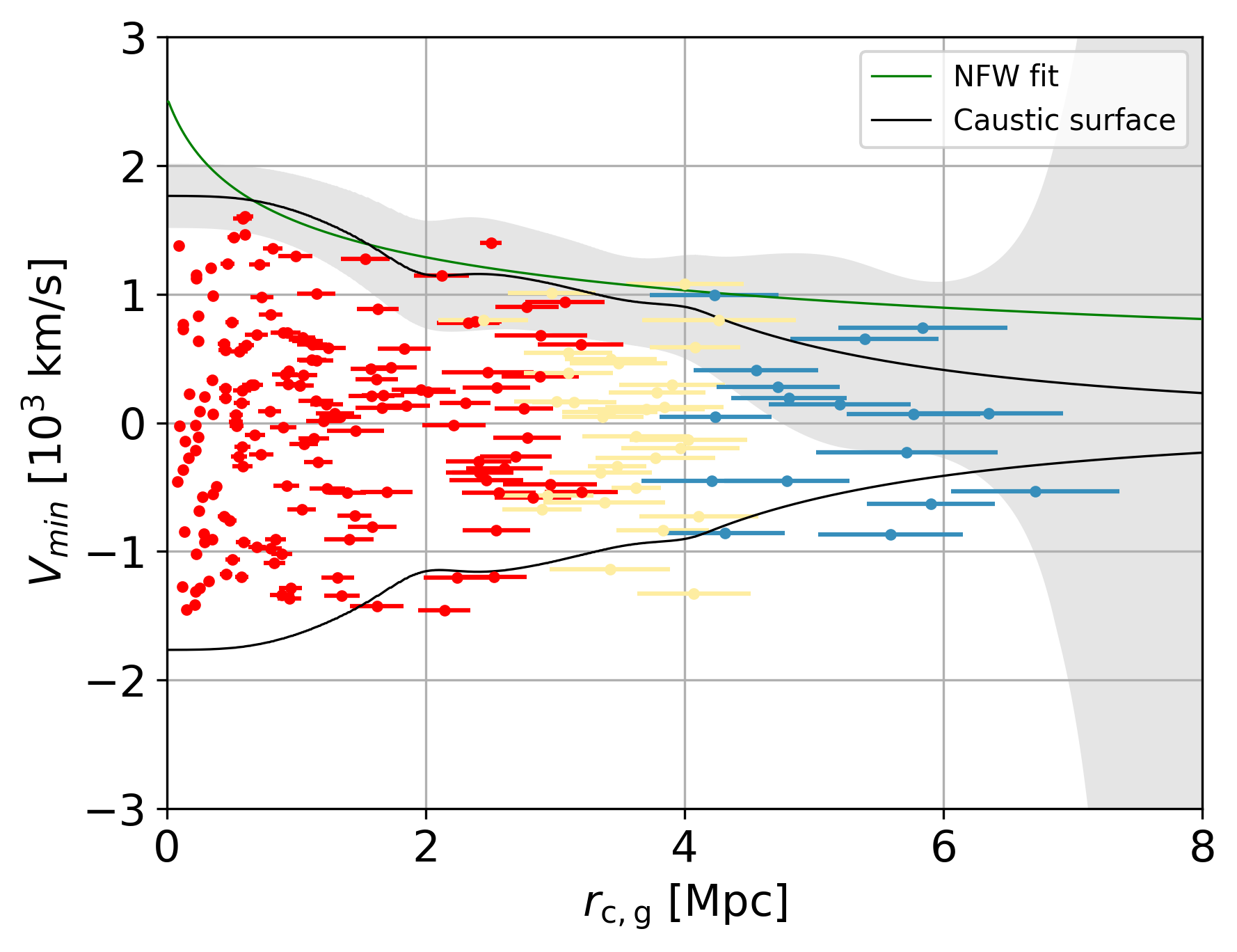}
\caption{\textit{Same as Fig.~\ref{fig:caustic_surface_full}, but the caustic surface is estimated using the 212 Coma members with CF4 distances to calculate their major infall velocity, Eq.~\eqref{eq:v_maj} (left) and their minor infall velocity, Eq.~\eqref{eq:v_min} (right).}} 
\label{fig:caustic_surface}
\end{figure}

Fig.~\ref{fig:caustic_surface} shows the same plots for the Coma member galaxies with CF4 distances and infall velocities detailed in Section~\ref{sec:radial_velocity}. 
The sparsity of this selection particularly at radii larger than the Coma core clearly causes the uncertainties in the caustics to increase compared to the complete dataset and diverge around $r_\mathrm{c,g} \approx 6$~Mpc. 

We then proceed to estimate the model-independent mass of Coma using the caustic technique. 
Theoretically, caustics can be drawn out to an arbitrary distance, therefore, we cut them when a mass corresponding to $M_{200}$ is reached.
Since the method is agnostic about the dynamical status of the cluster, we denote this mass as $\Mf^{\rm cau}$ and its corresponding radius as $r^{\rm cau}_{200}$ to distinguish these estimates from those obtained with other methods. 

Table~\ref{tab:mass_cons} lists the constraints on the mass alongside the projected approximation of $r^{\rm cau}_{200}$, Eq.~\eqref{eq:approx_dist}, for the complete dataset, and the one restricted to having CF4 distances for the different estimators of the infall velocities as detailed in Section~\ref{sec:radial_velocity}. 
For the complete Coma dataset including the field galaxies, we find $\Mf^{\rm cau} = (7.73 \pm 2.32) \times \Msun$, which is in good agreement with the estimates of $\Mf^{\rm cau} = (6.38 \pm 0.9) \times \Msun$ by \cite{Rines_2016}, being about $\sim 10\%$ higher. Once the field galaxies are excluded we find the mass to be $\Mf^{\rm cau} = (5.48 \pm 1.71) \times \Msun$. Reanalysing the dataset of \cite{Rines_2016} in the same way as our datasets, we also arrive at a caustic mass consistent with the previous value within their error bounds.
It is biased low with respect to $\Mf^{\rm cau} = (9.2 \pm 3.0) \times \Msun$ reported in \cite{bib:Sohn2017}, yet completely consistent within the estimated error bounds. Using the major and minor infall models we estimate the corresponding caustic masses to be $\Mf^{\rm cau} \sim 5.11 \times \Msun$ and $\Mf^{\rm cau} \sim 5.20 \times \Msun$, respectively. 
These estimates are completely consistent with the mass estimate from the complete dataset while the uncertainty on the caustic surface increases drastically towards the outskirts. 
Yet, the latter estimates are also fully compatible with the phase space taken from \cite{Rines_2016}. While being based only on approximately 20\% of the galaxy members with respect to our complete dataset and the one from \cite{Rines_2016}, we also find comparable relative uncertainties on the mass estimates. 

Comparing to other approaches to determine $r_{200}$, $r^{\rm cau}_{200}$ are biased low in general. 
Choosing a larger filling factor can alleviate this discrepancy, as the literature on this choice is not agreeing on one value that yields masses and radii for all possible cosmic structures that are consistent to those determined by other approaches, e.g. the virial theorem (see Section~\ref{sec:virial_theorem}). 

To estimate the NFW-masses for comparison, we fix the concentration to $c_{200} = 5$, \cite{Gifford:2013ufa, bib:Ettori2018}.
This is consistent with the expectation from the mass-concentration relation, see, for instance, \cite{bib:Merten_2015,bib:Dutton_2014, bib:Child_2018}, owing to the fact that estimating $c_{200}$ in addition requires more precise data in the inner regions of the cluster $r\sim 1~\mbox{Mpc}$. 
Fixing $c_{200}$, however, does not bias our mass estimation at $r_{200}$ and beyond.
Using this prerequisite, a NFW-fit to the caustic surface shows an equivalent trend for the mass estimates for major and minor infall datasets as $\Mf^{\rm cau, NFW} \in \left[ 4.55, 4.60\right] \times \Msun$, respectively. 
The mass estimates by the NFW-model fit, being model-dependent, are more robust and therefore have a slightly lower level of uncertainty than the model-independent caustic technique.

\begin{figure}
    \centering
    \includegraphics[scale = 0.65]{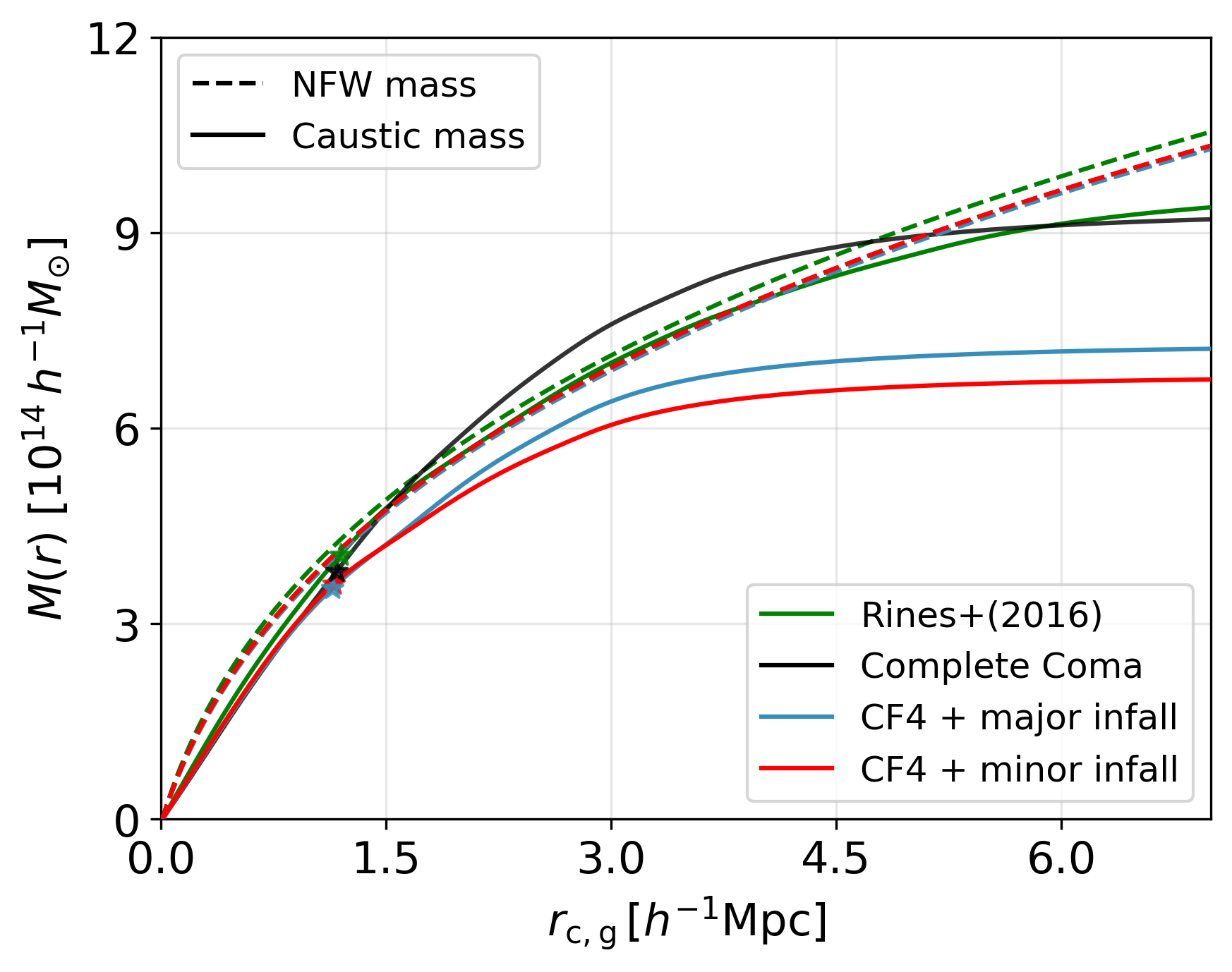}
    \caption{\it{Mass profiles of Coma estimated for the datasets listed in Table~\ref{tab:mass_cons} using the model-independent caustics method (solid lines) or a NFW-fit to the caustics (dashed lines). The $\star$ symbol along each mass profile depicts the $\Mf^{\rm cau}$ at the corresponding $r^{\rm cau}_{200}$.}}
    \label{fig:mass_profiles}
\end{figure}

As all mass estimates reported here have their own $r_{200}$, we contrast the mass profiles derived from each dataset in Fig.~\ref{fig:mass_profiles}. 
We find extremely good consistency with the earlier dataset of \cite{Rines_2016} for our major and minor infall models.
However, an under-prediction of the mass at $r_{\rm c, g} > 4\,{\rm Mpc}$ is observed for the latter two. 
This is consistent with the anticipation that the major and minor infall datasets possibly have less than $\sim 60\%$ completeness in these radial ranges. 
In contrast, the entire Coma set of 1092 galaxies is in very good agreement with the reconstruction of \cite{Rines_2016}, so that it predicts a mildly larger mass at smaller radii despite the fewer number of member galaxies. The reason for this is the higher completeness of our member galaxies in the inner regions ($r\lesssim 3~h^{-1} {\rm Mpc}$) possibly with the presence of a few interlopers, which are not discarded. Nevertheless, our approach is in extremely good agreement with the member selection in \cite{Rines_2016}.

Hence, we can conclude that the caustic technique allows us to estimate the mass profile to the outer regions of the cluster very well. 
However, the propagation of uncertainty through integration of the caustic surface increases the error on the mass profile drastically, unless extrapolated through an assumed mass model.

\begin{table}
\centering
\begin{tabular}{lccc}
\hline
\hline
Dataset & $n_{\rm gal}$ &$\Mf^{\rm cau} \, [\Msun]$ & $r^{\rm cau}_{200} \, [{\rm Mpc}]$\\
\hline
Rines+(2016) & $1147$ & $5.84 \pm 1.70$ & $1.71$\\
Complete Coma & $1092$ & $5.48 \pm 1.71$ & $1.68$\\
CF4 + Minor Infall & $212$ & $5.20 \pm 1.78$ & $1.65$\\
CF4 + Major Infall & $212$ & $5.11 \pm 1.92$ & $1.64$\\
\hline
\hline
\end{tabular}
\caption{Comparison of caustic mass estimates for different datasets, including the mass estimated from the dataset of \cite{Rines_2016}, Rines+(2016), processed in the same way as the other datasets for comparison. The first two datasets are processed with $H_0=70$~km/s/Mpc. Error bars on $\Mf^{\rm cau}$ are obtained by 1-$\sigma$ bounds on $M(r)$ at $r^{\rm cau}_{200}$.}
\label{tab:mass_cons}
\end{table}

\subsection{Virial Theorem}
\label{sec:virial_theorem}

Assuming that the ``Coma core'' data covers the virialised region, the virial mass of the core is calculated using the virial theorem, \cite{Limber:1960,bib:Bahcall1981,Heisler:1985,bib:Evans2011,bib:Tully2015,Benisty:2024tlv},  
\begin{equation}  
M_{\text{vir}} = \alpha \frac{\sigma_\mathrm{v}^2 r_\mathrm{G}}{G} \;, \quad \text{where} \quad r_\mathrm{G} = \frac{N} {\sum \limits_{i<j} 1/r_{ij}} \;.  
\label{eq:virial_mass}  
\end{equation}  
Here, $\sigma_\mathrm{v}$ is the velocity dispersion of galaxies in the core, $r_\mathrm{G}$ represents the harmonic mean of projected pairwise distances between the $N$ galaxies (weighted by their inverse separations $r_{ij}$), and $\alpha$ is a dimensionless geometrical factor that depends on the dark matter distribution and velocity anisotropy. For example, $\alpha = 3$ corresponds to isotropic orbits in an isothermal sphere, $\alpha = 2.6$ to isotropic velocities in an NFW-profile, and $\alpha = 2.4$ to specific anisotropy models. To account for uncertainties in the dark matter distribution, we adopt a broad uniform prior of $\alpha \in [2, 5.5]$.  

The velocity dispersions for galaxies in the virial core, derived from the minor and major infall models, as detailed in Section~\ref{sec:radial_velocity}, are 
\begin{align}  
\sigma_{\text{min}} &= \left( 752 \pm 42 \right) \times 10^{3}~\text{km/s} \;, \nonumber \\  
\sigma_{\text{maj}} &= \left( 821 \pm 46 \right) \times 10^{3}~\text{km/s} \;.  
\end{align}  
As shown in \cite{bib:Wagner2025}, the major infall model overestimates the true radial velocity dispersion, while the minor infall model underestimates it. Consequently, the true dispersion lies between these values.  

\begin{figure}
\centering
\includegraphics[scale=0.42]{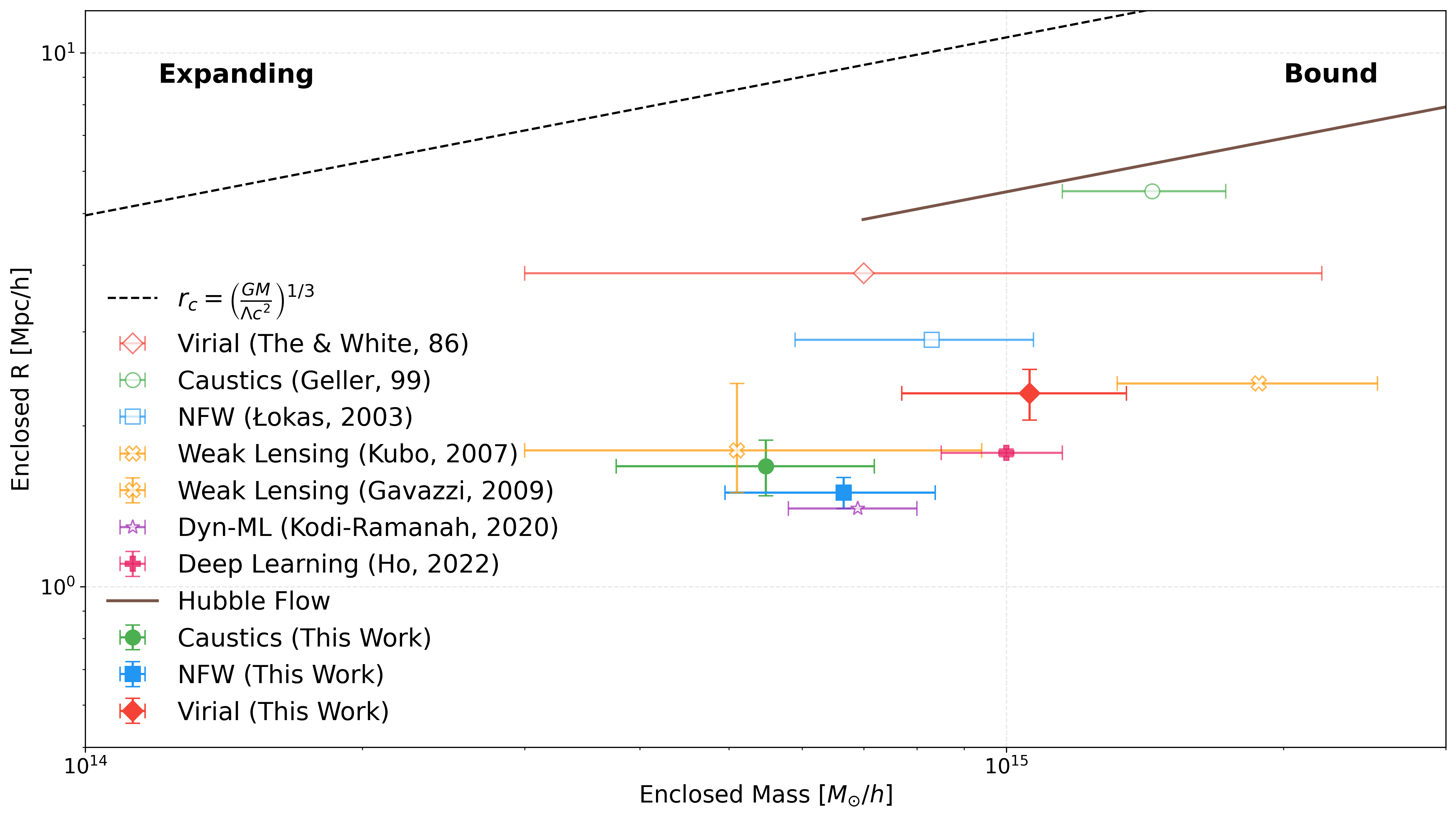}
\caption{\it{Comparison of mass estimates for Coma derived from results in this work and historical estimates from the literature with respect to the radius that encloses this mass: Our results (filled markers, colour-coded by method) include caustics, NFW-profile fits to caustics, CF4-based Hubble-flow, and CF4-based virial mass estimates (see Section~\ref{sec:mass_det}). Historical estimates (unfilled markers, colour-coded by method) are labeled with their respective references, \cite{bib:SinWhite1986,Geller:1999ci,bib:Hughes1989,Lokas:2003ks,Kubo:2007wt,bib:Gavazzi2009,Falco:2013bgy,KodiRamanah:2020nbc,Ho:2022lru,bib:Cha2025,bib:Costa2025}, Dyn-ML = dynamical machine learning. All masses are in units of $h^{-1}\, M_{\odot}$ for an $H_0$-independent, observation-based comparison. All of the mass estimates are inside the maximal turnaround radius possible as shown in \cite{bib:Pavlidou2014}.}} 
\label{fig:mass_com}  
\end{figure}

The weighted average projected distance $r_\mathrm{G}$ is approximated using Eq.~(\ref{eq:approx_dist}), which incorporates a logarithmic correction term $0.1 \ln 10 (\mu_i+\mu_j)$ to improve error estimates:  
\begin{eqnarray}  
r_\mathrm{G} &\approx& \frac{r_\mathrm{c} N}{k} \; \quad \text{with} \quad k \equiv \sum_{i<j} \frac{1 - 0.1 \ln 10 (\mu_i+\mu_j)}{\sin (\theta_{i,j}/2)} 
=  (2.26 \pm 0.06)~h^{-1}\,\text{Mpc} \;,  
\end{eqnarray}  
where $\theta_{i,j}$ is the angular separation between {the pairs of} galaxies in the core. For the CF4 survey calibration, $r_\mathrm{G} = (3.15 \pm 0.01)~\text{Mpc}$. Substituting these results into Eq.~\eqref{eq:virial_mass}, the virial mass for the minor infall model is  
\begin{equation}  
M_{\text{vir}}^{\text{min}} = \left( 1.06 \pm 0.29 \right) \times 10^{15}\,h^{-1} \,M_{\odot}\;,  
\end{equation}  
or $(1.45 \pm 0.15) \times 10^{15}~M_{\odot}$ using CF4 distances. For the major infall model, we find  
\begin{equation}  
M_{\text{vir}}^{\text{maj}} = \left( 1.26 \pm 0.34 \right) \times 10^{15}\,h^{-1} \,M_{\odot} \;,  
\end{equation}  
corresponding to $(1.72 \pm 0.21) \times 10^{15}~M_{\odot}$ with CF4 distances. As demonstrated in \cite{bib:Wagner2025}, the minor and major infall models enclose the true radial velocity dispersion, implying the virial mass lies between these estimates. Combining the results, we conclude the virial mass of the core is in the range between $(1.45 \pm 0.15) \times 10^{15}~M_{\odot}$ to $(1.72 \pm 0.21) \times 10^{15}~M_{\odot}$ for CF4 distances. This estimate is consistent with independent measurements of the Coma mass (see also Fig.~\ref{fig:mass_com}), for instance, from weak lensing studies in the SDSS~\cite{Okabe:2010}, which report $M_{200} \approx 1.3 \times 10^{15}~M_{\odot}$, further validating our results.  

\section{Discussion}
\label{sec:dis}

Gravitationally bound systems are subject to an interplay of the gravitational attraction by their constituents and the repulsion due to the expansion of the Universe. 
The transition between these two regimes has been investigated, for instance, for the Fornax cluster of galaxies \cite{Nasonova:2011md} and for the Virgo cluster \cite{bib:Sorce2016,bib:Kim2020,bib:DelPopolo2022} in our cosmic neighbourhood. Even though the Coma cluster is at a larger distance of about 100~Mpc, there have been many works analysing its central, virialised region and the {filaments in} which Coma is embedded.
Yet, there is hardly any investigation of the transition region, which is why we identify and characterise this region for Coma in this work. 

To identify the Coma cluster structure from the core to its Hubble flow with the least amount of necessary model assumptions, we develop a new processing pipeline from data selection, cluster member identification, to the required transformations for the cosmic distances and total velocities into the reference frame of the Coma cluster.

{First, to ensure homogeneity and consistency within the data used, we base our analysis on two datasets only: the SDSS~DR17 catalogue of galaxies equipped with spectroscopic redshifts and line-of-sight helio-centric velocities, and the Cosmicflows-4 catalogue containing independently measured cosmic distances. 
When cross-matching the datasets for the Coma region, we account for the varying value of the Hubble constant for different probes of cosmic distances by using the jointly calibrated distances.} 

Next, we establish a cluster member selection independent of a cosmological model (see Section~\ref{sec:member_selection} for details).
To do so, we collect all galaxies from SDSS~DR17 within a $10^\circ$ radius on the sky around the known Coma centre out to $z=0.05$. Then, we select all galaxies in $z$-direction that lie around the peak of the Coma centre out to the nearest local minima. 
Subsequently, we perform a density-based clustering (\texttt{dbscan}) of these galaxies on the sky and systematically test a broad range of reasonable parameters for it. 
We find that this clustering approach is able to give an estimate of the virial radius ($1.6^\circ\approx 1.95~h^{-1}~\mathrm{Mpc}$) and a lower limit of the turnaround radius ($3.99^\circ\approx 4.87~h^{-1}~\mbox{Mpc}$). 
The former is in agreement with prior estimates and the latter is a consistent upper bound on them (see Fig.~\ref{fig:mass_com}). 
In total, we identify 1092 cluster member galaxies separated into three regions within Coma, a virialised ``core'', an extended ``full'' set of Coma members, and the ``Coma outskirts'' of galaxies close to our turnaround radius. 
Using the highly precise redshift data and assuming that the centre of Coma has a vanishing peculiar velocity, we determine its line-of-sight velocity as $v_\mathrm{c} = 6996~\mathrm{km}/\mbox{s}$ and its cosmic distance $69.959~h^{-1}~\mathrm{Mpc}$.
As tested, all member galaxies lie within a small cone around the Coma centre, so that the small angle approximation is valid to determine their individual distances. 

Comparing our cluster member selection with the one from \cite{Rines_2016}, we find a high degree of overlap, 91\%, with the main difference that the selection by \cite{Rines_2016} extends beyond our turnaround radius. 
Both member selections are similarly compatible with a rough determination of cluster members by the caustic method, given the accuracy of this approach. 
Moreover, our member selection is similarly efficient in removing field galaxies and interlopers than conventional approaches which are mostly based on a Friends-of-Friends clustering followed by an interloper removal according to \cite{Hartog1996}. 

After cross-matching with CF4, only SDSS-selected 212 member galaxies and 479 field galaxies can be equipped with independent distances, which changes the statistical properties compared to the full member set. 
For these 212 members, the cluster centre lies at the CF4 distance modulus $\mu_\mathrm{c} = 34.89 \pm 0.03$, translating into $r_\mathrm{c} = 95.7~\mathrm{Mpc}$. 
Due to the small spread of the member galaxy $\mu$s around $\mu_\mathrm{c}$, we are able to Taylor expand the distances around $r_\mathrm{c}$, which also works for a limited amount of field galaxies. 
Subsequently, we set up the minor and major infall models to describe the radial velocities of all galaxies within Coma and in the field around it. 
The uncertainties in the distances and the velocities of these carefully selected galaxies are then greatly reduced because we anchor our Taylor expansion in the very precise mean value of $r_\mathrm{c}$.
The error propagation then includes a term linear in the distance-moduli differences between $\mu_\mathrm{c}$ and the $\mu$s of the galaxies. 

As detailed in \cite{bib:Tully2023}, combining different probes for cosmic distances in CF4 is a challenge because each probe is based on a different physical mechanism and synchronising them with respect to each other in a self-consistent way is not straightforward. 
This is why we choose to use CF4, providing synchronised distances, but we briefly comment on the spread of cosmic distances based on different probes and the spread in the inferred $H_0$-values in Section~\ref{sec:consistency}:
Combining the $v_\mathrm{c}$ inferred from SDSS~DR17 with $r_\mathrm{c}$ from the CF4-$\mu_\mathrm{c}$ yields $H_0 = (73 \pm 7)~\mathrm{km}/\mbox{s}/\mbox{Mpc}$.
Yet, probe-specific distances as shown in Fig.~\ref{fig:CF4_probes} lead to a spread in $H_0$ ranging from less than 72~km/s/Mpc to over 80~km/s/Mpc. 
Analogously, the spread of distances to the centre of Coma based on the individual, not-synchronised, {individual} CF4 probes ranges from less than 88~Mpc to more than 97~Mpc. 
These discrepancies, driven by sparse sampling and calibration biases, underscore the difficulty in achieving percent-level $H_0$-precision without addressing multi-probe systematics. 

{Based on this selection of galaxies, we can identify galaxies at distances around 10-20~Mpc from $r_\mathrm{c}$ that belong to the Hubble flow around Coma (see Fig.~\ref{fig:vec_dis_rel}). 
Trying to apply a Hubble-line fit to the velocity-distance relation, we find that the lack of galaxies in the Hubble flow at distances closer to $r_\mathrm{c}$ renders the fit unable to constrain the mass of Coma or the Hubble constant.
Calculating the velocity dispersion of these galaxies for the major and minor infall models, we can corroborate the findings made in \cite{bib:Karachentsev2006} for the M81-group that the velocity dispersion of these galaxies increases when we apply the major infall model compared to the minor infall model. 
As demonstrated in \cite{bib:Wagner2025} in terms of analytical derivations and $N$-body simulations, this is caused by the definition of the major infall model as a ratio of relative velocities by relative projected distances.}

In contrast to the inapplicable Hubble-flow fit, we assemble other mass estimates for the Coma cluster based on the caustic method and the virial theorem, as well as an upper limit from the turnaround radius.
{All these mass estimates} are consistent with each other and consistent with mass estimates from prior work.
All details can be found in Section~\ref{sec:mass_det} and a summary of all mass constraints for the respective radii is plotted in Fig.~\ref{fig:mass_com}. 
Compared to prior work shown, \cite{bib:SinWhite1986,Geller:1999ci,bib:Hughes1989,Lokas:2003ks,Kubo:2007wt,bib:Gavazzi2009,Falco:2013bgy,KodiRamanah:2020nbc,Ho:2022lru,bib:Cha2025,bib:Costa2025,Ho:2022lru}, which employs methods used in this work, weak lensing but also most recent artificial intelligence approaches, we can conclude that the spread in the mass-radius diagram is similarly broad than the spread in the distances to Coma in Fig.~\ref{fig:CF4_probes} for the same reason: every estimator is based on a different physical principle and they may even focus on effects at different distance scales from the centre. 

Extending the analysis of the cosmic expansion from the Hubble flow to the effect of $\Lambda$ in bound structures, \cite{bib:Pavlidou2014} state that the maximum value for the turnaround radius is $r_{\Lambda} = \sqrt[3]{G M/{\Omega_\Lambda H_0^2 }}$. This upper bound is shown in Fig.~\ref{fig:mass_com} as a dashed black line. All of our mass estimates and the ones from the literature are deeply inside this bounded area. Other galaxy groups also show the same behavior, for instance, the Local Group~\cite{Karachentsev:2008st,Benisty:2023clf,Benisty:2023vbz}. 

The DESI collaboration will greatly advance our understanding of the dark matter content in the Coma cluster by providing a much larger set of precise velocities and independently determined distances, vastly improving the sampling in the outskirts and infall regions~\cite{bib:Said2024}. This dataset will refine key dynamical parameters, like the velocity anisotropy profile and enabling a direct reconstruction of the escape velocity profile for the caustic method. These advancements will allow us to distinguish between different dark matter density profiles and to probe anomalies that could signal novel physics, such as self-interacting dark matter or modified gravity.  

\acknowledgments
We cordially thank Adi Zitrin for helpful discussions on the SDSS catalog, Nathan Secrest for his support on defining completeness and his insights into SDSS target-selection, {Brent Tully and Asta Heinesen for clarifications on the error budget of the recession velocity}. DB is supported by a Minerva Fellowship of the Minerva Stiftung Gesellschaft f\"ur die Forschung mbH and by a Short Term Scientific Missions (STSM) in Italy funded by the COST Action CA21136. DB acknowledges the contribution of the COST Actions CA21136 and CA23130. BSH is supported by the INFN INDARK grant and acknowledges support from the COSMOS project of the Italian Space Agency (cosmosnet.it).

\bibliographystyle{apsrev4-1}
\bibliography{ref}


\end{document}